\documentclass{aa}  

\usepackage[breaklinks=true,colorlinks,citecolor=blue]{hyperref}
\usepackage{graphicx,animate}
\usepackage{enumerate}
\usepackage{epsfig}
\usepackage{color}
\usepackage{txfonts}
\usepackage{natbib}
\usepackage{multirow}
\usepackage{ulem} 
\usepackage{amssymb}

\usepackage{movie15}

\usepackage{booktabs}
\usepackage{array}   

\usepackage{academicons}

\newcommand{\orcid}[1]{\href{https://orcid.org/#1}{\textcolor[HTML]{A6CE39}{\aiOrcid}}}
\usepackage{orcidlink}

\usepackage{xcolor}

\defcitealias{Mapping-model}{Paper I}
\defcitealias{Mapping-disk}{Paper II}
\defcitealias{Mapping-bulge}{Paper III}
\defcitealias{Mapping-extragalactic}{Paper IV}
\defcitealias{Mapping-sfh}{Paper V}

\definecolor{ao}{rgb}{0.0, 0.5, 0.0}

\newcommand{\kmps}{\rm km~s\ensuremath{^{-1} }\,}
\newcommand{\kmskpc}{km~s\ensuremath{^{-1}}~kpc\ensuremath{^{-1} }\,}

\newcommand{\Gaia}{{\it Gaia}\,}

\newcommand{\Lz}{\ensuremath{\rm L_z}\,}
\newcommand{\ELz}{\ensuremath{\rm E-L_z}\,}

\newcommand{\FeH}{\ensuremath{\rm [Fe/H]}\,}

\newcommand{\E}{\ensuremath{\rm E}}

\newcommand{\sz}{\ensuremath{\rm \sigma_{z}}}
\newcommand{\sphi}{\ensuremath{\rm \sigma_{\phi}}}
\newcommand{\sr}{\ensuremath{\rm \sigma_{R}}}

\newcommand{\vz}{\ensuremath{\rm V_z}}
\newcommand{\vp}{\ensuremath{\rm V_\phi}}
\newcommand{\vr}{\ensuremath{\rm V_R}}

\newcommand{\papername}{Rediscovering the Milky Way with orbit superposition approach and APOGEE data}

\newcommand{\thankssdss}{Funding for the Sloan Digital Sky Survey IV has been provided by the Alfred P. Sloan Foundation, the U.S. Department of Energy Office of Science, and the Participating Institutions. SDSS acknowledges support and resources from the Center for High-Performance Computing at the University of Utah. The SDSS web site is www.sdss4.org. \\

SDSS is managed by the Astrophysical Research Consortium for the Participating Institutions of the SDSS Collaboration including the Brazilian Participation Group, the Carnegie Institution for Science, Carnegie Mellon University, Center for Astrophysics | Harvard \& Smithsonian (CfA), the Chilean Participation Group, the French Participation Group, Instituto de Astrofísica de Canarias, The Johns Hopkins University, Kavli Institute for the Physics and Mathematics of the Universe (IPMU) / University of Tokyo, the Korean Participation Group, Lawrence Berkeley National Laboratory, Leibniz Institut für Astrophysik Potsdam (AIP), Max-Planck-Institut für Astronomie (MPIA Heidelberg), Max-Planck-Institut für Astrophysik (MPA Garching), Max-Planck-Institut für Extraterrestrische Physik (MPE), National Astronomical Observatories of China, New Mexico State University, New York University, University of Notre Dame, Observatório Nacional / MCTI, The Ohio State University, Pennsylvania State University, Shanghai Astronomical Observatory, United Kingdom Participation Group, Universidad Nacional Autónoma de México, University of Arizona, University of Colorado Boulder, University of Oxford, University of Portsmouth, University of Utah, University of Virginia, University of Washington, University of Wisconsin, Vanderbilt University, and Yale University.}

\newcommand{\thanksgaia}{This work presents results from the European Space Agency (ESA) space mission Gaia. Gaia data are being processed by the Gaia Data Processing and Analysis Consortium (DPAC). Funding for the DPAC is provided by national institutions, in particular the institutions participating in the Gaia Multi-Lateral Agreement (MLA). The Gaia mission website is https://www.cosmos.esa.int/gaia. The Gaia Archive website is http://archives.esac.esa.int/gaia.}

\newcommand{\thanksmiapb}{This research was supported by the Munich Institute for Astro-, Particle and BioPhysics (MIAPbP) which is funded by the Deutsche Forschungsgemeinschaft (DFG, German Research Foundation) under Germany´s Excellence Strategy-EXC-2094-390783311}

\newcommand{\thanksleiden}{The authors thank the organizers of the ``A new dawn of dwarf galaxy research'' workshop and the Lorentz Center~(Leiden) for the organizational and financial support of the meeting, which provided an insightful atmosphere which stimulated the progress of this work.}

\defcitealias{Mapping-1}{Paper I}
\defcitealias{Mapping-2}{Paper II}
\defcitealias{Mapping-3}{Paper III}
\defcitealias{Mapping-4}{Paper IV}
\defcitealias{Mapping-5}{Paper V}

\begin{document} 

\title{\papername\ I. Method validation} 

\titlerunning{Orbit superposition method in MW}
\authorrunning{S. Khoperskov et al.}

\author{Sergey Khoperskov$^1$\thanks{sergey.khoperskov@gmail.com}\orcidlink{0000-0003-2105-0763}, 
Glenn van de Ven$^2$\orcidlink{0000-0003-4546-7731}, 
Matthias Steinmetz$^1$\orcidlink{0000-0001-6516-7459}, 
Bridget Ratcliffe$^1$\orcidlink{0000-0003-1124-7378},
Ivan Minchev$^1$\orcidlink{0000-0002-5627-0355}, \\
Davor Krajnovic$^1$, 
Misha Haywood$^3$\orcidlink{0000-0003-0434-0400}, 
Paola Di Matteo$^3$, 
Nikolay Kacharov$^1$\orcidlink{0000-0002-6072-6669}, \\
Léa Marques$^{1,4}$,
Marica Valentini$^1$\orcidlink{0000-0003-0974-4148}, 
Roelof S. de Jong$^1$\orcidlink{0000-0001-6982-4081} }

\institute{$^1$ Leibniz-Institut für Astrophysik Potsdam (AIP),
              An der Sternwarte 16, 14482 Potsdam, Germany \\
              $^2$ Department of Astrophysics, University of Vienna, Türkenschanzstraße 17, A-1180 Vienna, Austria \\
              $^3$ GEPI, Observatoire de Paris, PSL Research University, CNRS, Place Jules Janssen, 92195 Meudon, France\\
              $^4$ Universität Potsdam, Institut für Physik und Astronomie, Karl-Liebknecht-Str. 24-25, 14476 Potsdam, Germany}

\abstract{We introduce a novel orbit superposition method designed to reconstruct the stellar density structure, kinematics, and chemical abundance distribution of the entire Milky Way by leveraging 6D phase-space information from its resolved stellar populations, limited by the spatial coverage of APOGEE DR17.}


\maketitle


\section{Introduction}

Our knowledge about the Milky Way~(MW) structure results from observations of individual stars. In 1785, William and Caroline Herschel made an extremely important discovery about the structure of our galaxy. By counting stars in the sky, they found that most of the stars they could see lay in a flattened structure, and the number of stars was about the same in any direction. They concluded that the star system to which the Sun belongs is in the shape of a disc or wheel, and the Sun must be near the middle of this structure~\citep{1785RSPT...75..213H}. Since then, the systematic counting of stars has been increasing continuously the number of known stars and improved our understanding of their host -- the MW galaxy~\citep[e.g.][]{1918ApJ....48..154S, 1920ApJ....52...23K, 1932ngcd.book.....A, 1979A&AS...38..423G, 1995gcts.book.....V, 1998A&AS..129..431H}.

Two centuries later, in 1989, the European Space Agency~(ESA) launched the Hipparcos spacecraft, which operated in orbit for 37 months, performing photometric and astrometric measurements and delivering positions, proper motions and parallaxes of the MW stars at the millisecond level of accuracy. The observations resulted in the Hipparcos catalogue of approximately $100$ thousand stars~\citep{1997A&A...323L..49P} which enabled the precise determination of the distance to the Galactic center~\citep{1998ApJ...494L.219P} and the constraint of key kinematic parameters of the solar neighbourhood, revealing the complex motions of the nearby stellar populations~\citep{1998MNRAS.298..387D, 1998AJ....115.2384D}. Auxiliary Tycho-1 and Tycho-2 catalogues~\citep{1997A&A...323L..57H, 2000A&A...355L..27H} have made it possible to study millions of stars in the MW with more precise proper motions and, thus, influenced many areas of modern astronomy~\citep{2009aaat.book.....P}. Nowadays, nearly a decade post-launch, it is hard to overstate the success of the still ongoing ESA space mission -- \Gaia~\citep{2016A&A...595A...1G, 2021A&A...649A...1G}. Perhaps the primary and the most fundamental outcome of this survey is our newfound ability to perceive the ensemble of the MW stars as a galaxy.

While the coverage by various stellar surveys continues to expand, reducing the 'Galaxia incognita'\footnote{The term 'terra incognita' employed in ancient cartography denotes regions that have not been mapped or documented. It is believed to have first appeared in Ptolemy's Geography around 150 AD.} within the MW, the number of stars with known positions, stellar parameters and kinematic information still peaks near the Sun, failing to reflect the underlying stellar density distribution. This discrepancy mostly stems from a fundamental limitation of our observations originating from the solar position within the MW disc midplane. Our location limits the volume of potentially observable stars, dependent on their apparent brightness, which is further influenced by interstellar medium~(ISM) extinction, source crowding in the dense inner regions of the Galaxy and so on. Although spectroscopic observations across different wavelength ranges, such as infra-red (e.g., APOGEE~\citep{2017AJ....154...94M} and soon MOONS~\citep{2020Msngr.180...18G}), partially mitigate the impact of dust, they do not fundamentally alter the distribution of observed stars as a function of distance from the Sun. In this case, the observations cover only a fraction of the selected area and thus might not be representative of the stellar population of this area in the sky. To probe the underlying stellar populations, one needs to know what fraction of stars was observed relative to the total number of stars in the same area and correct for possible selection biases or to demonstrate their absence~\citep{2018MNRAS.476.3278C, 2017A&A...606A..97N, 2017MNRAS.468.3368W}.

Different techniques have been introduced to correct for the observational biases, assuming certain completeness of the observed sample as a function of some parameters~\citep[see, e.g.][]{2014ApJ...790..127B, 2014ApJ...796...38N, 2014ApJ...793...51S, 2019A&A...621A..17M, 2021MNRAS.501.2954B, 2022MNRAS.510.4626B, 2023arXiv231018258W, 2023A&A...677A..37C}. For instance, \cite{2018MNRAS.476.3278C} used weighting schemes to correct for the metallicity bias introduced by the target selection in their sample of SEGUE main-sequence turn-off stars and G and K dwarf star~\citep[see also][]{2012ApJ...761..160S}. \cite{2017A&A...606A..97N} found no large selection function effect on the metallicity distribution function~(MDF) and the vertical metallicity gradient for APOGEE~(DR13, \citealt{2017ApJS..233...25A}), however, their sample is restricted from $7$ to $9$~kpc and within $2$ kpc from the midplane. \cite{2019A&A...621A..17M} showed that the effect of the selection function is more prominent for surveys that have a complex target allocation strategy and whose footprint is more patchy~(e.g. APOGEE and Gaia-ESO), while for surveys with contiguous footprints, such as GALAH, RAVE, and LAMOST, the selection function has almost no effect in the extended solar vicinity~\citep{2017MNRAS.468.3368W, 2018MNRAS.476.3278C}. 

Generally speaking, various sorts of analyses of the MW stellar populations rely on relatively small sub-samples of stars drawn from the spectroscopic catalogues by applying user-defined cuts motivated by the scientific problem and the desired precision of stellar parameters. Hence, the selection function of that sample of stars is always unique and thus requires particular treatment; otherwise, the recovered kinematics and stellar abundance variations may not be representative of particular regions (or components) in the MW. Therefore, there is still a strong need for a selection function independent analysis of data from ongoing and future spectroscopic surveys~\citep{2013A&ARv..21...61R}.

Over the last years, the availability of large amounts and high-precision observational data, particularly from recent and ongoing ground-based spectroscopic surveys~(e.g., Gaia-ESO~\citep{2022A&A...666A.121R}, RAVE~\citep{2020AJ....160...83S}, LAMOST~\citep{2012RAA....12..723Z}, APOGEE~\citep{2017AJ....154...94M}, GALAH~\citep{2021MNRAS.506..150B}) and space mission \Gaia~\citep[ESA][]{2016A&A...595A...1G,2023A&A...674A...1G}, has revolutionised our understanding of the Galactic stellar populations. Various datasets offer a wealth of information that can be leveraged to reconstruct the MW present-day structure~\citep{2018MNRAS.481L..21P, 2018Natur.561..360A,  2020A&A...634L...8K, 2020MNRAS.494.4291C, 2022A&A...658A..91A} and uncover some episodes of its assembly history~\citep{2018MNRAS.478..611B,2018ApJ...863..113H, 2018Natur.563...85H, 2019MNRAS.486.3180K, 2020ARA&A..58..205H}. In parallel, using this information, multiple models of the MW have been developed, which one can divide into ones tailored to constrain the formation and enrichment history of the Galaxy~\citep[see, e.g.][]{2009MNRAS.396..203S,2021MNRAS.507.5882S,2019ApJ...884...99F,2023MNRAS.523.2126P} and models aimed to recover its present-day structure~\citep[see, e.g.][]{2001ApJ...556..181D,2003A&A...409..523R,2012MNRAS.426.1328B,2014A&A...564A.102C,2016ApJ...830...97T,2021ApJ...910...17P,2024MNRAS.527.1915B,2023MNRAS.520.1832B}. However, understanding the MW as a galaxy remains a formidable challenge with its complex interplay of stars, gas, and dark matter, encoding the underlying structure and kinematics of the Galaxy. 

Outside the MW, significant progress has been made in developing sophisticated techniques to probe the dynamics and structure of external galaxies. One such method that has attracted significant attention is the Schwarzschild orbit superposition technique~\citep[][see also ~\cite{1984A&A...141..171P,1988ApJ...327...82R,1997ApJ...488..702R,1998ApJ...493..613V,2004MNRAS.353..391T,2004ApJ...602...66V,2006MNRAS.366.1126C,2008MNRAS.385..647V,2020ApJ...889...39V,2020ascl.soft11007J,2020MNRAS.496.1579Z,2021MNRAS.500.1437N}]{1979ApJ...232..236S}. This powerful approach offers a unique way to model the gravitational potential and kinematics of galaxies by superposing a library of stellar orbits. The application of the Schwarzschild method has seen remarkable success in modelling the dynamics of external galaxies, providing insights into their dark matter content~\citep[see, e.g.][]{1997ApJ...488..702R,2010ApJ...719.1481V,2011MNRAS.415..545T,2023A&A...675A.143C,2024MNRAS.530.4474L}, bar structures~\citep[see, e.g.][]{2013MNRAS.435.3437W,2015MNRAS.450.2842V,2020IAUS..353..176V, 2021MNRAS.500..838P,2022ApJ...941..109T, 2023arXiv231000497T, 2024arXiv241005374K}, central black holes~\citep[see, e.g.][]{1998AJ....116.2220V,2002MNRAS.335..517V,2003ApJ...583...92G,2006MNRAS.367....2H, 2009MNRAS.399.1839K, 2013ARA&A..51..511K, 2016ApJ...818...47S, 2016ApJ...831..134V}, assembly history~\citep[see, e.g.][]{2023A&A...672A..84D,2021A&A...647A.145P,2022A&A...660A..20Z} and orbital structure~\citep{2005MNRAS.357.1113K, 2008MNRAS.385..647V,2018MNRAS.473.3000Z, 2021MNRAS.508.4786D, 2024MNRAS.534..861T}.

In this paper, we present a novel orbit superposition approach based on the APOGEE-like mock observations of a simulated MW-like barred galaxy. Such an approach has not been adapted in the context of the MW so far, as it presents specific challenges, but also new opportunities for a deeper exploration of observational data very much needed in light of the coming new generation of spectroscopic surveys, like 4MOST~\citep{2019Msngr.175....3D}, SDSS-V~\citep{2023ApJS..267...44A}, MOONS~\citep{2020Msngr.180...18G} and WEAVE~\citep{2023MNRAS.tmp..715J}. Our method follows the original orbit superposition approach first introduced by \cite{1979ApJ...232..236S}, where a library of stellar orbits was used to make a model of a triaxial elliptical galaxy without constraining the solution by kinematic information. We demonstrate that under reasonable assumptions regarding the galactic potential, this method makes it possible to correct the APOGEE-like mock data for the selection function and even more to go beyond a specific survey spatial footprint and recover the unbiased kinematics and abundance information across the entire galaxy. In the follow-up works, we apply this technique to reconstruct the three-dimensional stellar density distribution, kinematics and chemical abundance composition of different components of the MW galaxy. 

The structure of the paper is as follows. In Section~\ref{sec::method_method} we explain in detail the methodology of the orbit superposition method in the context of the MW-like data and construction of the mock APOGEE-like data. In Section~\ref{sec::model_results}, we present the results of the orbit superposition using two mock catalogues based on different selections of star particles, test the reconstructed kinematics of stars and the ability of the method to reproduce the stellar metallicity distribution. We discuss the limitations and future prospects of our approach in Section~\ref{sec::method_discussion}. The summary of the paper is given in Section~\ref{sec::method_summary}.

\section{Methodology}\label{sec::method_method}
The objective of a classic Schwarzschild orbit superposition method, in the context of external galaxies, is to find the parameters of gravitational potential (DM and baryons contributions) and to recover the phase-space structure of stellar components constrained by the observed distribution of stars (imaging) and line-of-sight kinematic information. Solving this problem requires finding a consistent solution to both the collisionless Boltzmann equation and the Poisson equation. This is usually accomplished by combining some basis orbits~(or library of orbits) with corresponding weights to reproduce a density (or light) distribution that aligns with the potential used to generate the orbits. The total galactic potential can be represented by a sum of contributions of different components~(dark matter, stars, gas, central black hole):
\begin{equation}
    \Psi_{tot}({\bf r}) = \Psi_{DM}({\bf r}) + \Psi_s({\bf r}) + \Psi_g({\bf r}) + \Psi_{BH}({\bf r})\,.
\end{equation}
While in external galaxies the total potential $\Psi_{tot}({\bf r})$ is unknown a priori, thanks to the amount and various sources of data, the total MW mass~\citep{2019MNRAS.487.2685E, 2020MNRAS.498.5574E, 2020MNRAS.494.4291C,  2021MNRAS.501.2279V}, the contribution of different components~\citep{2015ApJS..216...29B, 2017MNRAS.465...76M, 2017A&A...598A..66P, 2019ApJ...871..120E}, halo shape~\citep{2013ApJ...773L...4V,2019A&A...621A..56P,2021MNRAS.501.2279V} and even the 3D structure of the stellar disc, including the complex bar/bulge inner region~\citep{2013MNRAS.435.1874W,2015MNRAS.450.4050W,2017MNRAS.465.1621P, 2022MNRAS.514L...1S}, are constrained quite well. Therefore, if we assume that the total potential of the galaxy and, more specifically, its stellar component are known or well constrained by certain independent measurements, then the 3D stellar density distribution $\rho_s({\bf r})$ can be easily recovered from the Poisson equation:
\begin{equation}
    \nabla^2\Psi_s({\bf r})  = 4 \pi G \rho_s({\bf r}) \,.
\end{equation}
and only the complete phase-space distribution function remains unknown. In order to obtain the complete distribution function, one can use the orbits of stars calculated in the total potential of the galaxy, $\Psi_{tot}({\bf r})$. This can be done using separable triaxial potentials, where all orbits are regular, conserve three integrals of motion and can be calculated analytically~\citep{1985MNRAS.216..273D,1993ApJ...409..563S}. In practice, the orbital library is usually limited by several families of orbits~\citep{1987ApJ...321..113S,1996MNRAS.283..149Z,2008MNRAS.385..647V}: short-axis tubes, outer and inner long-axis tubes, box orbits and so on. In the case of the MW, once a certain potential is adopted, one can integrate orbits of real stars using positions and velocities available from \Gaia and ground-based spectroscopic surveys as the initial conditions. Therefore, the library of orbits for the orbit superposition modelling can be based on the orbits of real objects and include all sorts of orbital families governed by the potential and the input phase-space distribution function. 

\begin{figure}
    \centering
    \includegraphics[width=1\hsize]{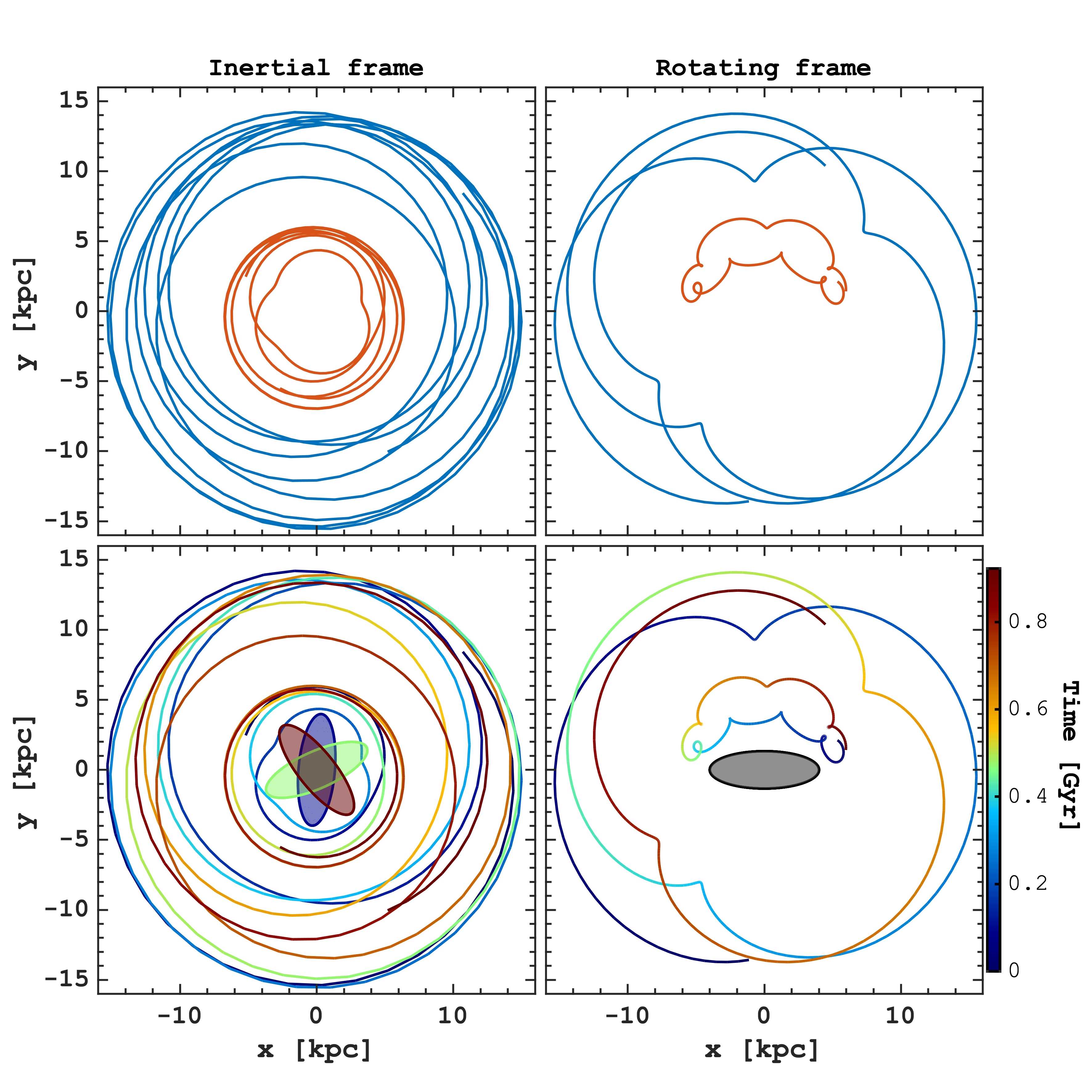}
    \caption{Example of two orbits~(red and blue) of disc stars in a barred potential in inertial~(left) and rotating with the bar~(right) reference frames. In the bottom panels, the orbits are colour-coded to indicate the progression of time. The ovals in the figure represent the bar orientation, with three different configurations colour-coded to denote changes in its orientation over time in the inertial frame. Notably, the orientation of the bar varies with time in the inertial reference frame while, by definition, it remains fixed in the rotating frame. Therefore, the orbits in the rotating frame allow us to exclude the time dependence of particle positions along the orbits and prevent phase-mixing, enabling us to use them in the orbit superposition approach.}
    \label{fig::example_orbits}
\end{figure}

In order to fill in the unknown~(or unobserved by a given Galactic survey) parts of the distribution function, following the canonical Schwarzschild modelling, one can solve the linear equation:
\begin{equation}
    \rho_s({\bf r}) = \sum^N_i w_i \rho({\bf r})_i\,\label{eq::model}
\end{equation}
where $w_i$ are the orbit weights, and each orbit has a 3D density distribution $\rho({\bf r})_i$, representing how much time a star spends in a given position. The solution of Eq.~\ref{eq::model} provides us with the weights $w_i$ for the orbit of each star in the observational sample whose superposition results in the adopted stellar density. 

It is crucial to emphasize the key distinctions between the orbit superposition approach we propose here for the MW and traditionally employed Schwarzschild models used to derive parameters of external galaxies. The primary divergence lies in the assumptions made regarding the knowledge of the gravitational potential, including individual contributions from dark matter, stars, and gas. For the MW, it is reasonable to assume that the total potential~($\Psi_{tot}({\bf r})$), as well as the contributions from stellar component~($\Psi_s({\bf r})$ or $\rho_s({\bf r})$), are known. Conversely, in the case of external galaxies, the gravitational potential is ab initio unknown, and its parameters are typically constrained using available kinematic and/or luminosity data. This can be achieved through an iterative fitting procedure, allowing for adjustments to the parameters of the potential~(disc and DM halo scales) and its functional form.

Therefore, the question we are seeking an answer to is not the parameters of the MW potential but rather its unbiased phase-space distribution function, which is normally obscured by the selection function of different surveys and their spatial incompleteness. In particular, the resulting orbit superposition solution makes it possible to map stellar kinematics beyond the survey footprint, essentially encompassing the entire galaxy. The resulting velocity distribution will remain unbiased against the survey selection function. If stellar parameters such as abundances and ages are available for stars in the sample used to construct the orbital library, it becomes possible to transfer these `labels' along the orbits and `paint' their distribution across the galaxy, which again will take into account the weights of populations with different stellar parameters, as they are evolutionarily linked to the orbits of stars. 

\begin{figure*}
    \centering
    \includegraphics[width=1\hsize]{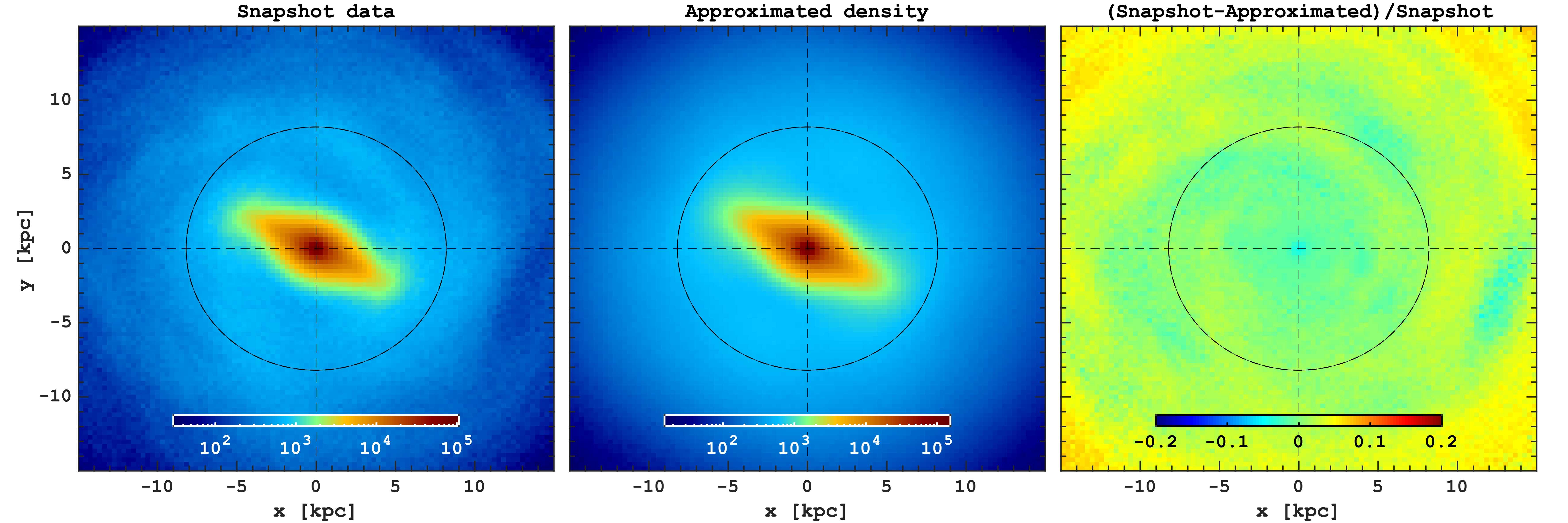}
    \caption{Approximation of the stellar disc density from the simulation. The surface density of star particles in a snapshot is shown in the left panel. In the middle panel, we present the surface density of the stellar disk obtained from the potential approximated using multipole expansion~(see Section~\ref{sec::method_simulation} for details). The rightmost panel shows the difference between the two maps, divided by the snapshot stellar density. Notably, the approximated density~(and potential) reproduces well the large-scale structure of the simulated galaxy, including the bar 3D structure; however, by construction, small-scale asymmetries such as spiral arms are not captured. The black circle shows the solar radius of $8.2$~kpc. The bar orientation is chosen to be $27^\circ$ relative to the horizontal axis, reproducing the MW bar positioning.}
    \label{fig::potential_sim}
\end{figure*}

\subsection{Simulation and potential reconstruction}\label{sec::method_simulation}

For the above-described method verification, we use a single snapshot from a simulation of an isolated MW-like barred galaxy from an $N$-body/hydrodynamical simulation, presented in \cite{2024MNRAS.528.3576V} and used to match the present-day MW kinematics. In this work, we do not re-scale the simulation to match the MW as we do not aim to reproduce any MW parameters but rather to validate the orbit superposition approach. The simulation also includes star formation, the basic chemical abundance cycle described in \cite{2021MNRAS.501.5176K} and \cite{2024A&A...690A.147H} from which we will use only metallicity information to test the goodness of our approach in the reconstruction of stellar parameters evolutionary-coupled with the orbits of stars. We, therefore, select a snapshot when the bar and the X-shaped bulge are already well developed, and the simulated galaxy looks visually similar to the MW~(see Fig.1 in \cite{2024MNRAS.528.3576V}). However, in order to highlight the limitations of our approach, we have deliberately chosen a snapshot with relatively strong spiral arms and not relaxed vertical motions of stars, which by construction can not be reproduced by Schwarzschild orbit superposition methods assuming a dynamical equilibrium. For the snapshot analysis, we orientated the bar by $27^\circ$ relative to the Sun-galactic centre line. 

The distinct feature of the MW stellar component is the bar, and in this work, we aim to include such a non-axisymmetric component in the orbit superposition modelling. In such a case, we will construct the orbital library in a rest frame rotating with the rotational frequency of the bar. This allows us to consider the positions of stars along the orbits time-independent because the bar position is always fixed to the present-day orientation relative to the Sun. We illustrate this coordinates transformation in Fig.~\ref{fig::example_orbits} where we show two orbits of random disc particles in a barred potential in the inertial~(left) and rotating~(right) frames. From the perspective of Galactic stellar populations, replacing individual stars with orbits implies that for each observed star, numerous other stars exist that are unobserved for various reasons yet follow identical orbits. That is why using a rotating rest frame is vital for `observing' these hypothetical stars at the same moment in time. It also allows us to avoid the phase-mixing of the distribution function naturally happening in an evolving potential, e.g., a rotating bar.

As the first step, we take particle positions and masses and create a smooth approximation of the galactic potential using the AGAMA code~\citep{2019MNRAS.482.1525V}. Since the simulation includes star particles of different ages and, thus, different kinematics, and we aim to reconstruct the potential of the bar and the X-shaped structure in as much detail as possible for the 3D stellar density~(potential) approximation, we use the spherical-harmonic expansion with $l_{max}=20$ and $m_{max}=20$. We recall that in such a case, we are unable to take into account spiral arms in the approximated version of the gravitational potential, which will result in minor but expected deviations of stellar kinematics and abundance distribution obtained in the orbital space compared to the original snapshot data. We approximate the potential of the simulated galaxy using three components: DM, stars and gas. Therefore, we can obtain the approximated 3D density distributions for each component separately. In Fig.~\ref{fig::potential_sim} we compare the face-on stellar surface density in the snapshot~(left) against its approximation with the spherical-harmonic expansion~(right). The residual map demonstrates a very good approximation of the stellar disc mass distribution, including the 3D structure of the inner region and the central bar.

\begin{figure*}
    \centering
    \includegraphics[width=1\hsize]{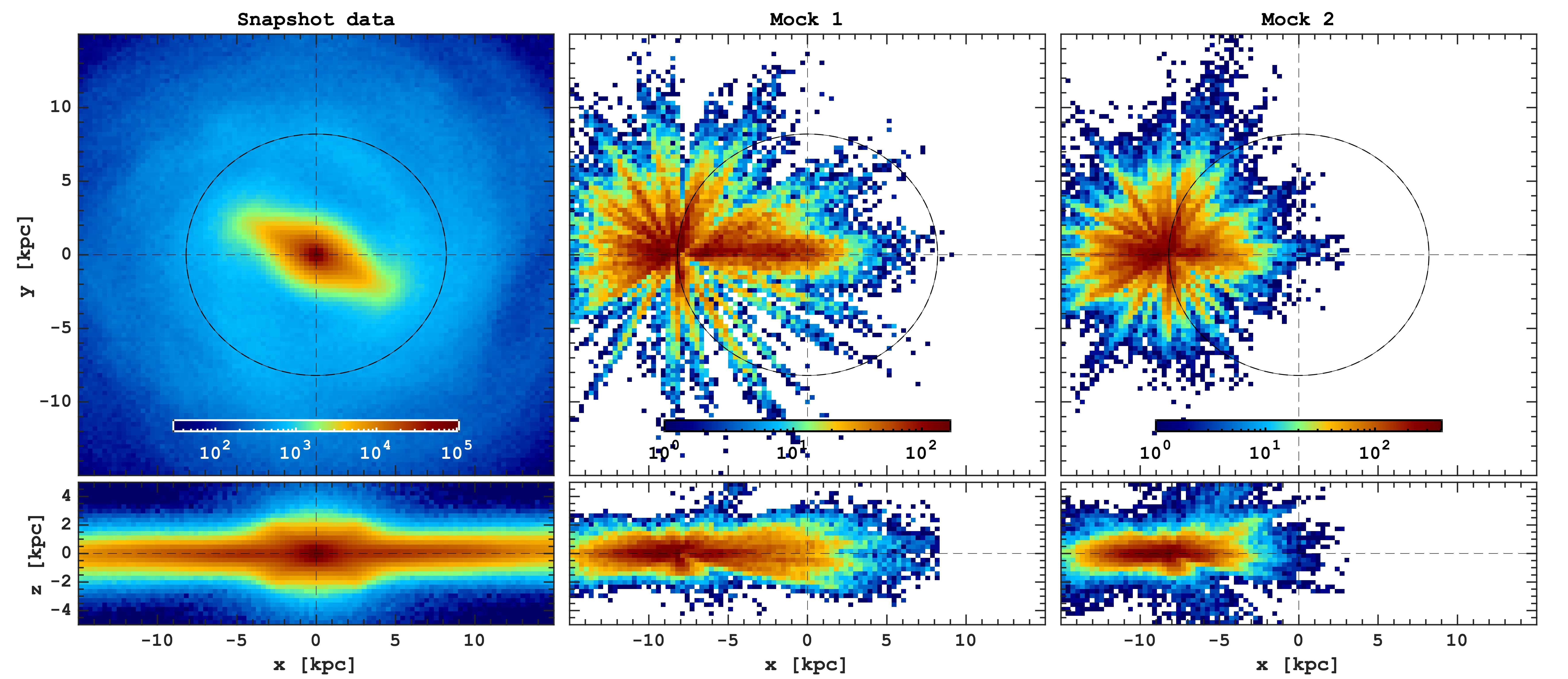}
    \caption{Initial selection of star particles from the simulation mimicking the APOGEE footprint. In the left panel, the surface density of all star particles in the snapshot is shown. The middle and right panels show two mock catalogues which qualitatively reproduce the APOGEE DR17 giants~(Mock 1) and red clump~(Mock 2) stars footprints, respectively~(see Sec.~\ref{sec::method_selection} for details). The number of star particles in the snapshot is 7.343.898, while the APOGEE-like selections include 40.354~(Mock 1) and 39.596~(Mock 2). In all the panels, the bar orientation is chosen to be $27^\circ$ relative to the horizontal axis, reproducing the MW bar positioning.}
    \label{fig::initial_selection}
\end{figure*}

\subsection{Selection of APOGEE-like samples from simulation}\label{sec::method_selection}

So far, we do not have access to the 6D phase-space information based on all stars across the MW, and the spatial selection function~(or distribution function) of large-scale disc surveys is quite complex. Therefore, it is vital for us to understand how much the results obtained using orbits of a limited sample of stars provide meaningful information about a larger area across the Galaxy. In the following, we use the snapshot simulations to extract particles, which will be used to make mock observations. The spatial distribution of these stars and their metallicities will be used as inputs for the orbit superposition models. The main purpose is to test how a selection bias (spatial footprint) affects the results of the modelling. We, therefore, mimic the APOGEE-like selection footprint by sampling the $N$-body simulation particles according to the distance distribution in the APOGEE catalogue in each HEALpix, assuming a HEALpix level of 4, which is a fairly low sky resolution, but it gives a reasonably good mock catalogue by design reproducing the spatial footprint of the MW survey. 

In order to test the effect of the survey footprint and, thus, different samples of orbits, we created two mock catalogues. The first one~(Mock 1), is based on the APOGEE giant stars selection, where we adopt $\rm log\ g<2.2$. Limiting to smaller $\rm log\ g$ values, the selection minimizes potential systematic uncertainties in abundance measurements, but the higher luminosity of these giant stars allows them to cover a larger area across the MW~\citep[see, e.g.][]{2022ApJ...928...23E, 2023A&A...676A.108C, 2023ApJ...954..124I, 2024A&A...690A.147H}. The second sample~(Mock~2) includes stars with $\rm 2.5 < log\ g < 3.6$, making it relatively local but still covering a substantial area around the Sun. This sample of the MW stars is often characterised by a relatively lower age determination uncertainty~\citep{2023A&A...678A.158A, 2023MNRAS.522.4577L}. The number of star particles in the APOGEE-like mock selections includes 40.354~(Mock~1) and 39.596~(Mock~2), while the total number of star particles in the snapshot is 7.343.898. Both mock samples are shown in Fig.~\ref{fig::initial_selection}, where the density peaks at the solar position and shows typical pencil-beam-like features. Below, we refer to the results of the orbit superposition modelling as Model 1 and Model 2, based on Mock 1 and Mock 2, respectively.

\section{Results}\label{sec::model_results}

\subsection{Orbits weight calculation}\label{sec::results_weights}
In Section~\ref{sec::method_method} we described a general idea behind the orbit superposition modelling in the context of the MW. In this section, we focus on more practical steps which we have developed using the simulated MW-like galaxy.

\begin{figure*}
    \centering
    \includegraphics[width=1\hsize]{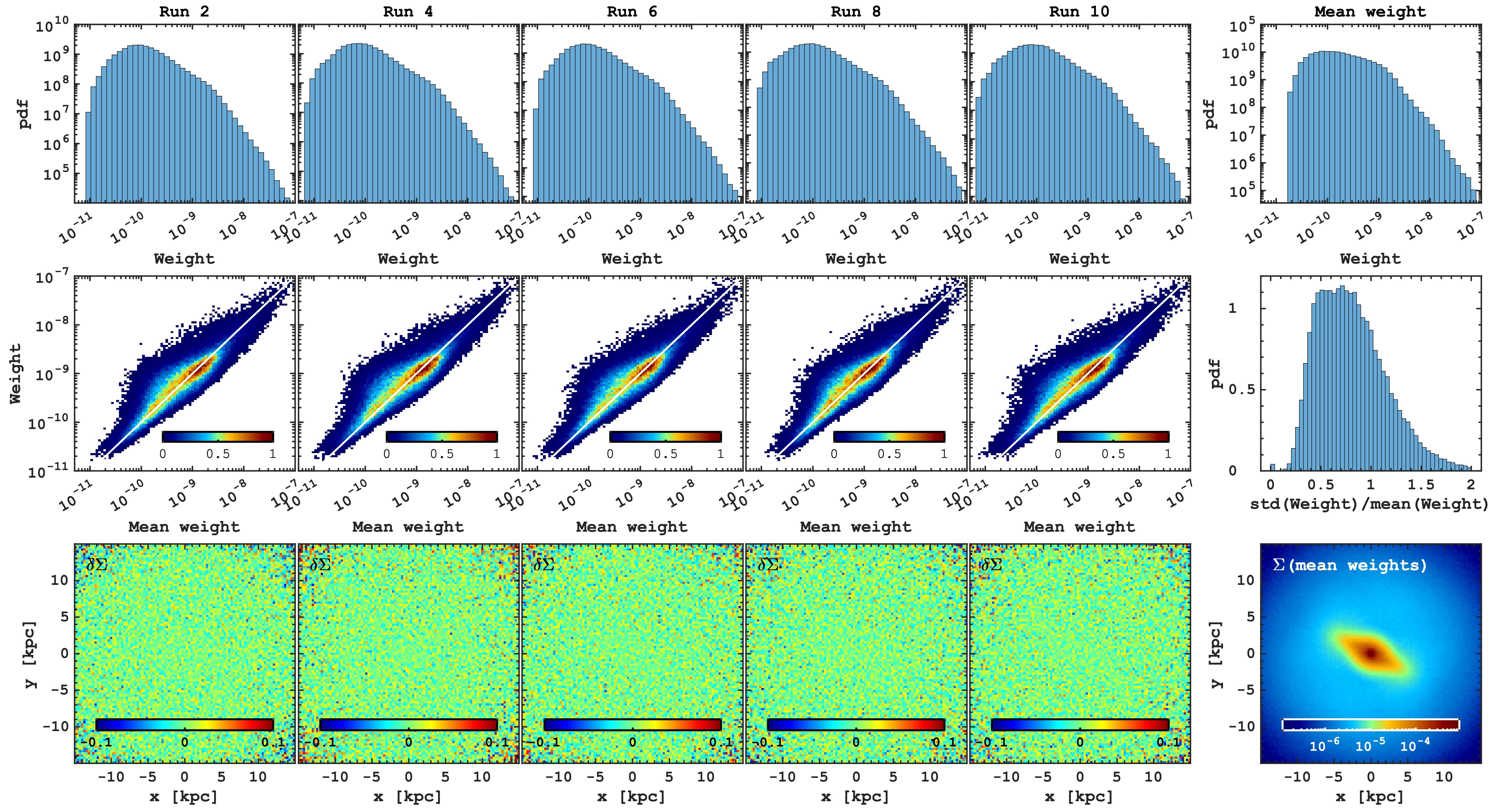}
    \caption{Analysis of weights of orbits across various orbit superposition model realizations using Mock 1. The five left panels of the top row illustrate the weight distribution in five realizations of the orbit superposition using different 10\% randomly sampled orbits, while the rightmost column shows the distribution of mean weights averaged across these realizations. In the second row, the five left panels reveal the relationship between mean weights and those obtained in different realizations of orbit superposition. The rightmost panel in this row presents the ratio between the standard deviation of weights and their mean. The bottom panels depict the relative difference between the stellar surface density obtained in various realizations and the surface density obtained using the mean orbital weights. Although a certain level of stochasticity in the distribution of the weights is seen, all the realizations successfully recover the same density distribution, which is nearly identical to the density based on the mean weights. The observed discrepancy in the orbital weight distribution across different realizations is likely attributed to a non-orthogonal orbital library. We, therefore, recommend using weights averaged across several orbit superposition realizations for more stable results.}
    \label{fig::weights_analysis}
\end{figure*}

We use the approximated 3D stellar density distribution of the galactic stellar disc component, which calculation is detailed in Sec.~\ref{sec::method_simulation}. The stellar density distribution~($\rho_s({\bf r})$) is then binned onto a 3D cube of 30 kpc in size~(from $-15$ kpc to $15$ kpc), with 50 bins along each direction. This Cartesian grid is also employed to determine the contribution of individual orbits~($\rho({\bf r})_i$).

The orbits of stars with initial conditions from both Mock~1 and Mock~2 were integrated over $5$~Gyr in a reference frame rotating at a constant angular speed corresponding to the measured bar pattern speed of $22$~\kmskpc. We use an instantaneous value of the bar pattern speed measured directly using the bar orientation in the previous and subsequent snapshots. Next, each orbit was discretized into 500 data points~(3 coordinates and 3 velocities) equally cadenced in time, which were then translated into density values on the 3D Cartesian grid mentioned above. Furthermore, to ensure the axial symmetry of the reconstructed density, each orbit was mirrored along every axis. We have experimented with other possible symmetries and did not find any substantial differences~\citep[see the detailed exploration of the orbit mirroring on the Schwarzschild modelling solution in][]{2022A&A...667A..51T}; thus, we stick to the simple mirroring of the orbits which also reduces computational costs. We note that the mirroring of each orbit does not imply that we assume all orbits to be symmetric relative to the principal axes. In fact, it means that for each asymmetric orbit, there are corresponding mirrored counterparts in our library.

We, therefore, have constructed the discrete versions of $\rho_s({\bf r})$ and $\rho({\bf r})_i$ where ${\bf r} = \{x_j,y_j,z_j\}_{j=1,...,50^3}$. Generally speaking, a solution of Eq.~\ref{eq::model} is unique if the number of variables $w_i$ matches the number of the discrete density values $\rho_s({\bf r})$, and the basis functions~(library of orbits) represent an orthogonal basis~\citep[see discussion in ][]{1979ApJ...232..236S}. However, our approach does not guarantee these conditions, potentially leading to the degeneracy of the solution~\citep{1996MNRAS.283..149Z}. To address this issue, for a given orbital library, we calculate non-negative weights $w_i$ using the least square minimization for 10\% randomly~(without repetitions) selected subset of orbits~(we experimented in 2-30\% range and find 10\% to be optimal in terms of solution stability and computational efforts), ensuring that every subset fits well the entire 3D stellar density $\rho_s({\bf r})$. This procedure has been repeated ten times for the whole sample of orbits, so we obtained ten realizations of weights for each orbit. Then, in order to minimise a possible degeneracy of the solution, we use the mean values of weights across these different realizations for further analysis. In Fig.~\ref{fig::weights_analysis}, we present the variation of weights across different realizations for Model~1, displaying the weight distribution for five out of ten realizations~(top row) and the mean weights~(top right). The weight distributions exhibit considerable width, spanning nearly four orders of magnitude, but the distributions closely resemble each other across the various realizations. Upon comparing the mean weights with those obtained in individual realizations, we observe a close one-to-one relation~(middle row of Fig.~\ref{fig::weights_analysis}); however, some scatter remains noticeable. Analysis of the ratio between the mean weights and the standard deviation across ten realizations reveals that individual measurements typically do not deviate by more than a factor of 2-3~(middle row, right panel). Nevertheless, the stellar density reconstructed using the individual weights is identical to that obtained using the mean weights (bottom panels). Therefore, we conclude that employing the averaged weights while helping to mitigate the degeneracy of the orbit superposition solution results in the desired solution for the disc stellar density.

\begin{figure*}
    \centering
    \includegraphics[width=1\hsize]{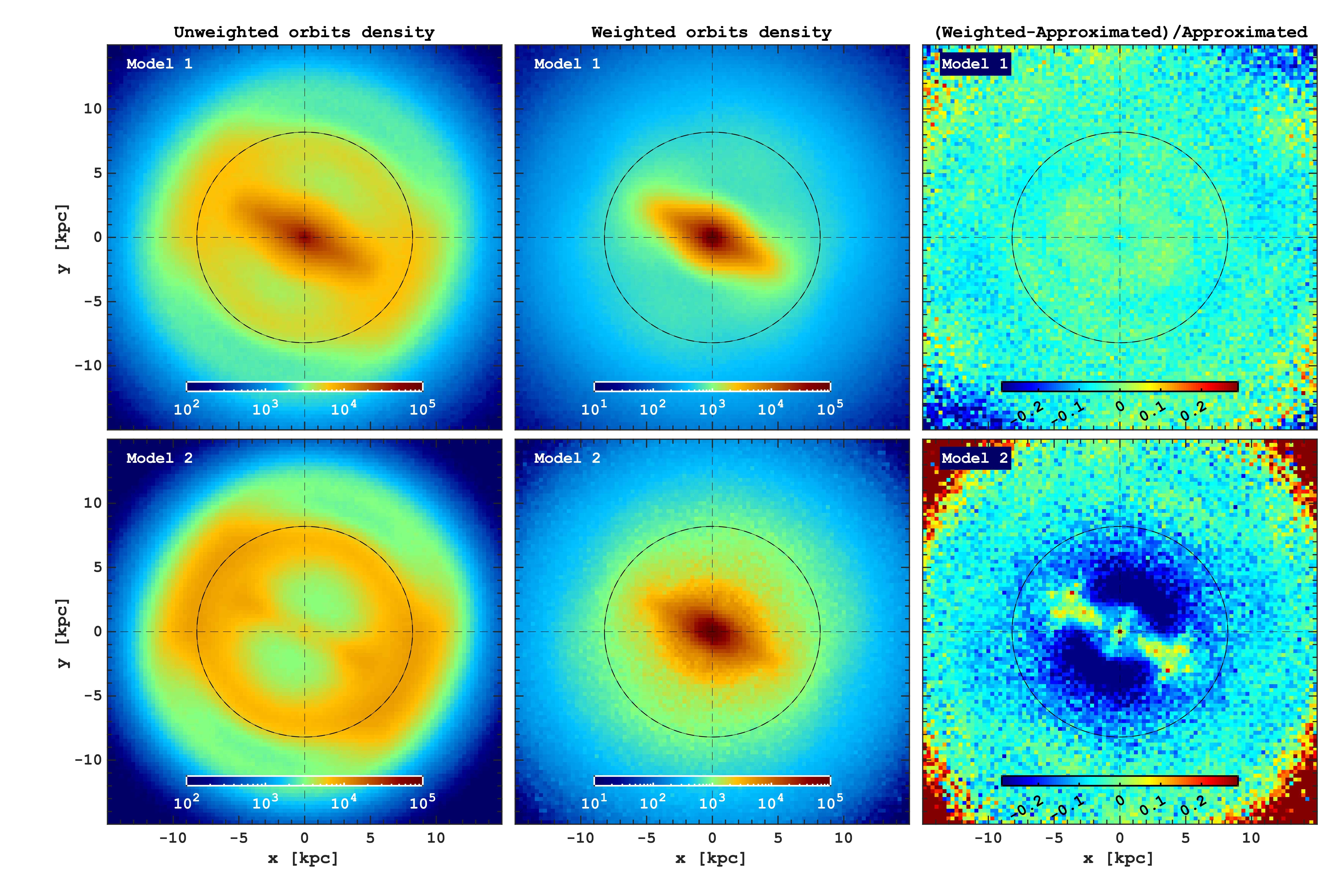}
    \caption{Results of the orbit superposition method. The left panels display the stellar density of orbits with constant weights, essentially presenting the stacked library of orbits of all star particles from Mock~1~(top) and Mock~2~(bottom). The middle panels depict the stellar density reconstructed using weighted orbit superposition where the weights are calculated as the mean across ten orbit superposition realizations (see details in Sec.~\ref{sec::results_weights}). The rightmost panels illustrate the relative residuals between the orbit superposition reconstruction and gravitational potential-based approximation (see Fig.~\ref{fig::potential_sim} and Sec.~\ref{sec::method_simulation} for more details).}
    \label{fig::density_reconstruction_maps}
\end{figure*}

\begin{figure}
    \centering
    \includegraphics[width=1\hsize]{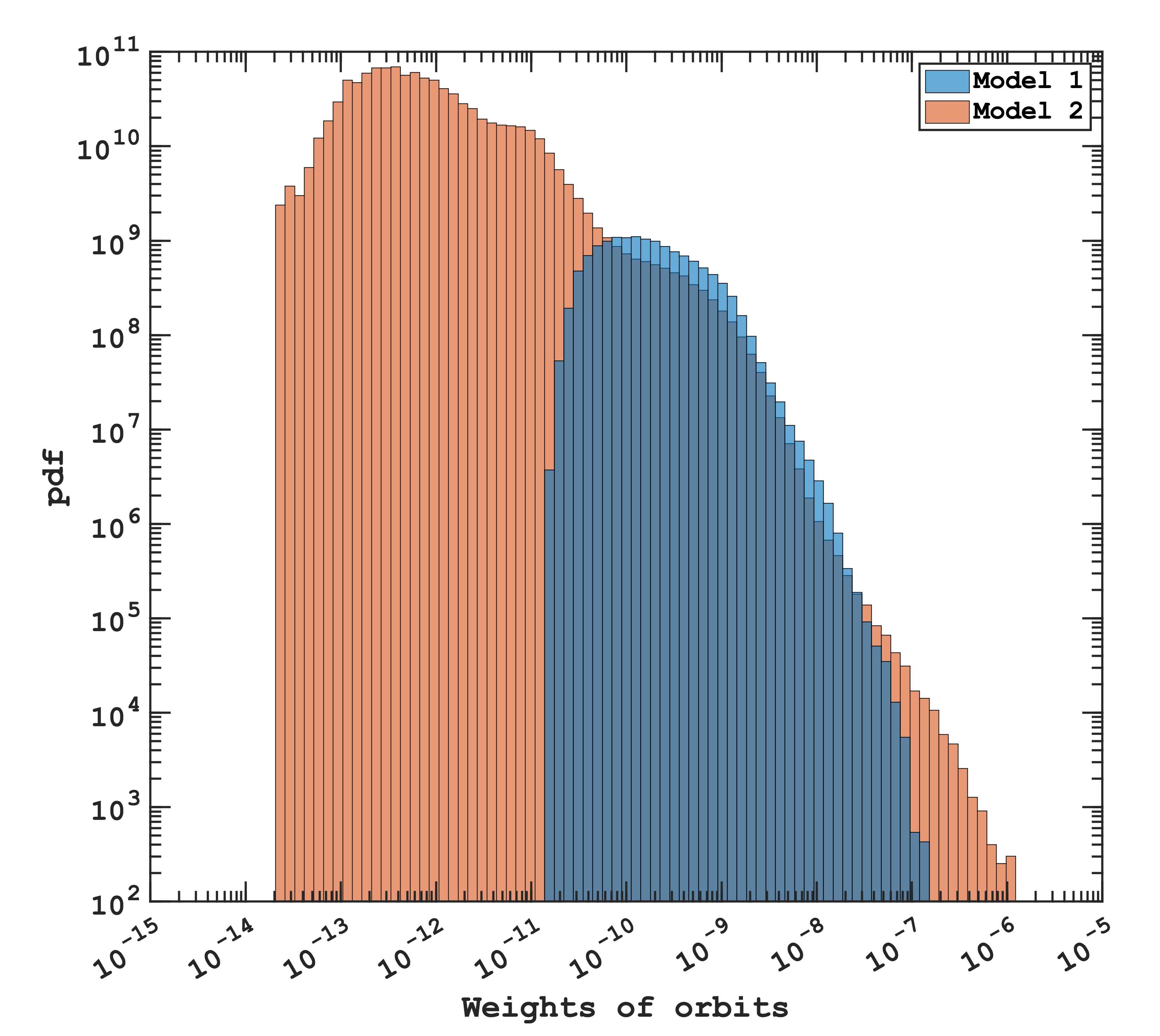}
    \caption{Mean weights distribution in two orbit superposition models obtained in ten realizations for the APOGEE-like Mock 1~(blue) and Mock 2~(blue; see details in Sec.~\ref{sec::method_selection}). The weight values are given in relative units. The broad distribution of weights in Model~2 indicates a worse convergence of the solution compared to Model 1.}
    \label{fig::weights_dist}
\end{figure}

\begin{figure*}
    \centering
    \includegraphics[width=1\hsize]{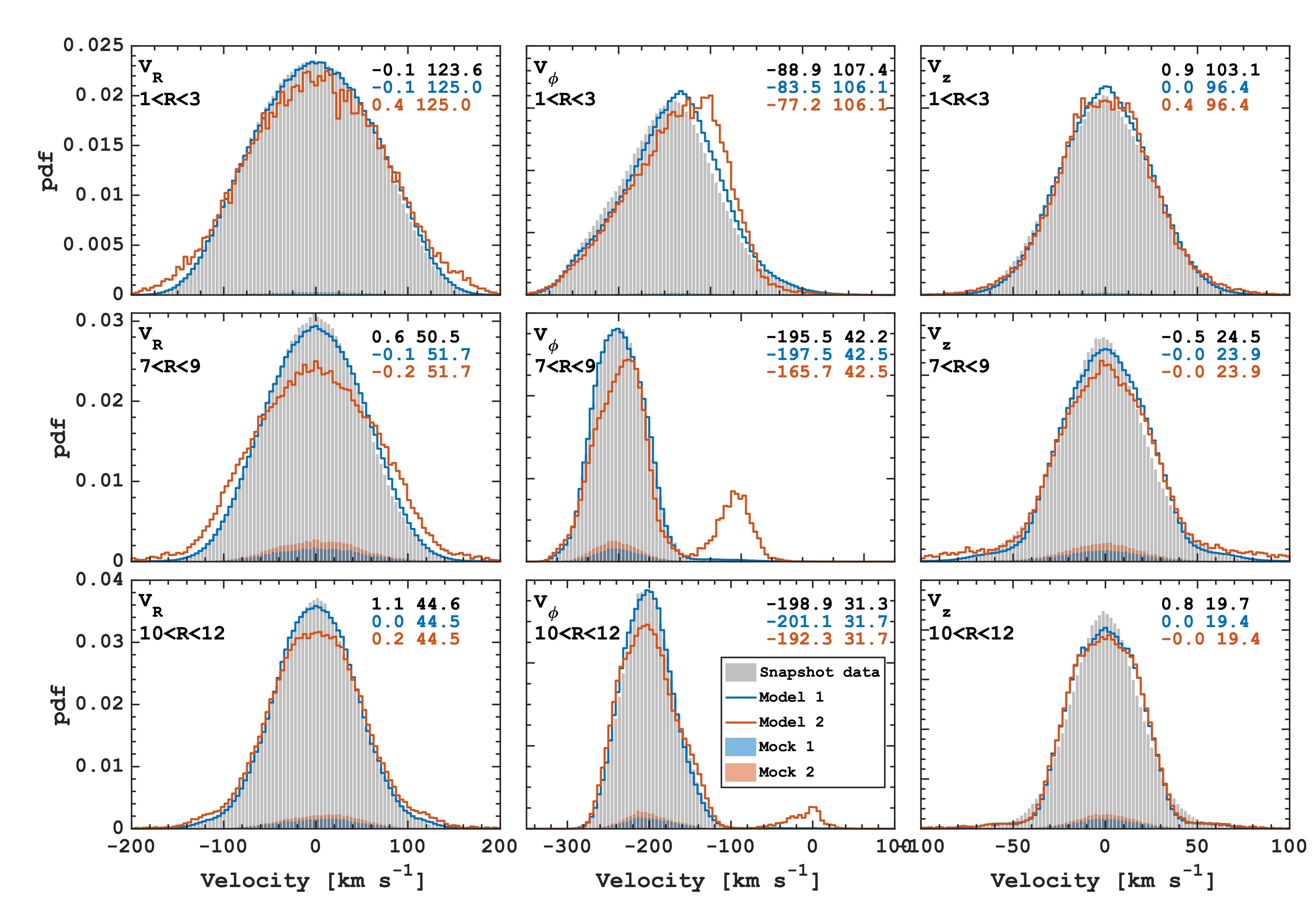}
    \caption{Distribution of velocity components~(\vr\ in left, \vp\ in the middle and \vz\ in the right) in the galactic rest frame within three 2~kpc-wide annuli centred at 1~(top), 8~(middle), and 11~kpc~(bottom). The snapshot data are represented by the grey-filled areas, while the results of the orbit superposition based on Mock 1 and Mock 2 are depicted in blue and red colours, respectively. The blue and red-filled areas correspond to the initial mock data. For each panel, the numbers correspond to the mean and standard deviation of the velocity distribution in \kmps. 
    We emphasize that the orbit superposition method we employ aims to recover the approximated stellar density distribution; the figure shows that Model~1 reproduces the kinematic information very well. At the same time, this approach is unable to capture spiral arms or the non-equilibrium kinematic features of the disc and bar. Furthermore, the effectiveness of the orbit superposition technique is contingent upon the accuracy of the potential approximation.}
    \label{fig::velocity_distributions}
\end{figure*}

\subsection{Stellar density reconstruction}\label{sec::results_density}

The results of the stellar density reconstruction are depicted in Fig.~\ref{fig::density_reconstruction_maps}. In the left panel, we present a stacked distribution of all orbits obtained in our models ($\displaystyle \sum^N_i \rho({\rm r})_i$). Notably, a striking difference emerges between our mock APOGEE-like selections. Model~1 yields an orbital density peaking at the galactic centre, indicating that our sample includes a significant number of stars with apocenters within our initial selection. Consequently, when mapping the orbits of these stars, they naturally contribute to the inner galaxy. Just beyond the bar, there appears to be an artificial (as it does not appear in the snapshot data) ring-like overdensity, likely populated by stars on nearly circular orbits and dominated at the solar radius. Although the model based on Mock~2 exhibits a similar orbital density, the bar-like overdensity is less pronounced. This difference is evidently due to the mock selection being more tightly confined to the solar radius, with less extension towards the centre of the galaxy. This result agrees with the finding of \cite{2022A&A...659A..80W}, who used the unweighted orbit superposition and found a similar inner ring-like structure in the MW between the planar bar and corotation.

The middle panels of Fig.~\ref{fig::density_reconstruction_maps} display the best solutions for the recovered stellar density, or, in other words, the weighted density distribution of all orbits in each model. As seen, the unweighted orbits based on Mock 1 and 2 are not identical~(see left panels), as their sample is controlled not only by the gravitational potential but also by the initial distribution function defining the orbital library in each model. Here, the distinction between the orbit superposition solutions for Model~1 and Model~2 is quite evident. Model~1 exhibits a striking resemblance to the approximated stellar density, a similarity further emphasized in the residual density map depicted in the rightmost panel where the relative difference between the two densities is less than $5\%$. For Model~2, although the overall density structure looks like a barred galaxy, the recovered density distribution is far from the one desired. In particular, the bar overdensity is less prominent than the approximated density, while the surrounding inner region is too dense and shows a prominent spheroid-like component in the centre. The quality of the stellar density reconstruction in Model~1 and Model~2 is somewhat reflected by the distribution of the mean weights, shown in Fig.~\ref{fig::weights_dist}. The best solution based on Mock 2 exhibits an extremely broad distribution of weights, spanning approximately nine orders of magnitude, compared to the more compact distribution observed in Model~1. This likely serves as an indicator of the degeneracy of the orbit superposition model, which can be helpful if the target density is weakly constrained. Therefore, we suggest that such a broad distribution of weights may signal a problematic solution alongside poor convergence of the stellar density solution derived~(see Fig.~\ref{fig::density_reconstruction_maps}) from the orbit superposition.

These best solutions for the stellar surface density in Fig.~\ref{fig::density_reconstruction_maps} demonstrate a rather obvious result that a superposition of not every random sample of stars is sufficient to recover the density structure of the galaxy. One can consider an extreme scenario where, for various reasons, the selection of stars includes only cold-orbit stars, and none of them passes through the Galactic centre. In terms of Schwarzschild modelling, this implies that the orbital library does not include the full range of orbital families in a given potential. However, it appears that present-day spectroscopic data from APOGEE DR 17~(Mock 1) alone are adequate to explore the entire MW using orbit superposition methods. Obviously, one can leverage datasets like \Gaia DR3, where the spatial coverage of stars with relatively precise 6D phase-space information extends even beyond the Galactic centre. If the high-precision chemical abundance information from larger samples of stars is needed, a promising approach might be to use the data from several spectroscopic surveys using homogenized spectroscopic datasets~\citep{2024A&A...690A..54T}.

At this point, one might consider disregarding the analysis of Model~2, given its failure to recover the stellar density precisely. However, since the ground truth about the real MW is unknown, it remains crucial to illustrate potential biases and their origin that could come out in the orbit superposition models for a large but obscured initial phase-space selection. Also, it is important to showcase situations in which the results obtained can still offer valuable information.

\subsection{Kinematics}\label{sec::results_kinematics}
In the following section, we analyze the stellar kinematics of the simulated galaxy recovered using the orbit superposition approach. It is important to note that, unlike traditional Schwarzschild methods, no kinematic information from our sample of stars has been used to calculate the weights of the orbits.

First, we examine the distribution of the orbit mass-weighted velocity components at different galactocentric radii and compare them to the ground truth~(snapshot data). In Fig.~\ref{fig::velocity_distributions}, we display the distributions of radial (\vr), azimuthal (\vp), and vertical velocities (\vz) in three circular two kpc-wide annuli centred at 2, 8, and 11 kpc. We recall that our orbit superposition approach aimed to recover the approximated stellar density distribution rather than the snapshot density distribution itself. Thus, despite a very good density approximation~(see Fig.~\ref{fig::potential_sim}), some differences between the snapshot data and the reconstructed stellar kinematics may be expected. 

The orbit superposition model based on Mock~1 reproduces the snapshot kinematics with great fidelity, and we observe only a few minor discrepancies. In particular, the radial velocity component distribution~(\vr, left column) almost perfectly recovered at all radii highlighted by the match between the blue lines~(Model 1) and grey shaded area~(snapshot data). At the same time, in the inner galaxy, the azimuthal velocity distributions~(\vp, top middle panel) do not perfectly align; it appears that the reconstructed velocity distribution is shifted towards lower absolute values, as evidenced by the difference in the mean velocity ($-83.5$~\kmps versus $-88.9$~\kmps in the snapshot data). Since we do not observe such a mismatch at larger radii~(middle and bottom panels of the middle row), it is likely that the approximated density, which we recover using orbit superposition, does not perfectly reproduce some small details~(lopsidedness or a weak misalignment relative principal axes) of the inner bar region. Additionally, there are slight discrepancies in the vertical velocity distribution, where the central peaks of the reconstructed velocity distributions are slightly lower. Nevertheless, these differences are hardly noticeable when comparing the mean and dispersion values, as indicated by the numbers in each panel.

While the mean and velocity dispersion values in Model 2~(red colour) remain very close to those of the snapshot data, as anticipated, it fails to capture multiple features of the corresponding distribution functions. Although the overall shapes of the radial and vertical velocity distributions (left and right columns) appear to be in qualitative agreement with the snapshot data, they do not reproduce the tails and central values of these distributions. A more striking artefact emerges in the azimuthal velocity distribution, which reveals a secondary component at low absolute velocity values. This kinematic feature, which either rotates slowly in the centre or remains non-rotating at larger radii, is naturally associated with a spheroid-like component, seen already in the stellar density solution in Fig.~\ref{fig::density_reconstruction_maps} (see Model 2 in the bottom middle panel). The emergence of such a feature in the orbit superposition solution is readily interpretable, as this method leverages existing star orbits to populate the 3D stellar density within a specified volume. To determine the stellar density solution at the centre, significant weight must be assigned to orbits that pass through the mock footprint. However, Mock 2 does not seem to capture a substantial number of stars near the galactic centre, relying only on hot radial orbits for the possible solution. These orbits possess minimal angular momentum, and when substantial mass is attributed to them, their cumulative effect manifests as an artificial spheroid with low net rotation. This underscores the critical importance of the initial star sample selection for the orbit superposition method, as it should allow the construction of a representative distribution function. Consequently, the failure of Model 2 is anticipated, yet it effectively demonstrates the machinery behind the orbit superposition approach.

\begin{figure*}
    \centering
    \includegraphics[width=1\hsize]{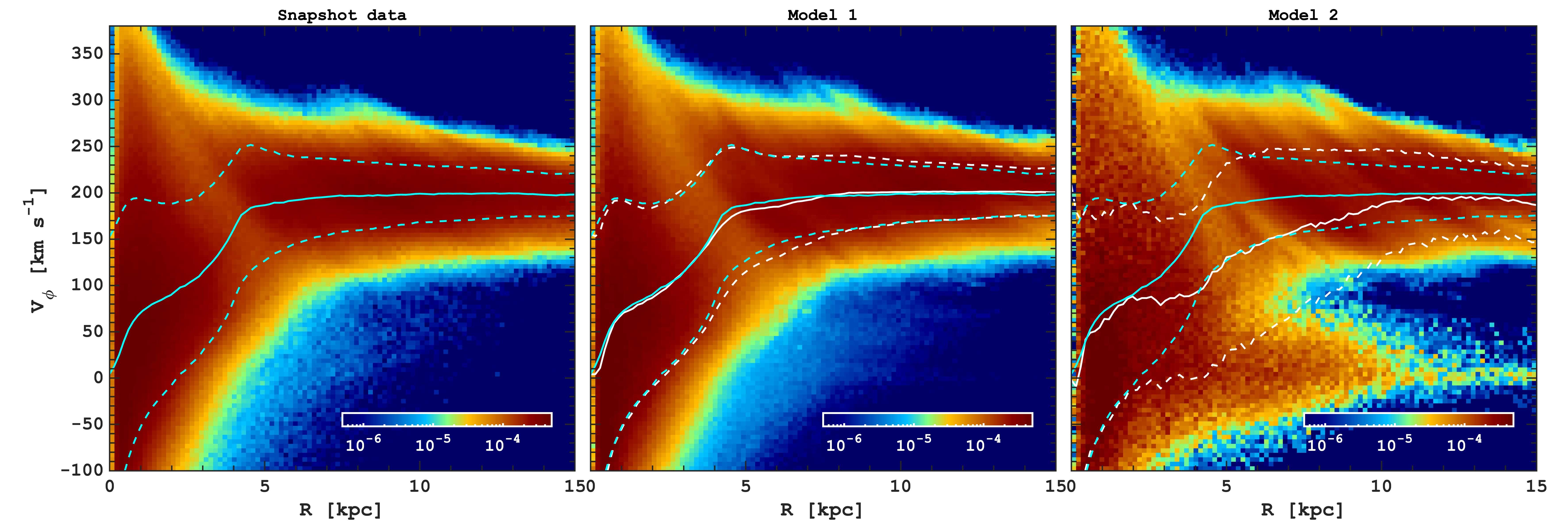}
    \caption{Reconstruction of the rotational velocity distribution as a function of the galactocentric distance. The left panel displays the stellar density distribution~(normalized by the total stellar mass) in the simulation, while the middle and right panels depict the stellar density reconstructed using orbit superposition based on Mock~1 and Mock~2, respectively. In all panels, the cyan lines represent the mean~(solid) and mean $\pm$ standard deviation~(dashed) of the snapshot data. Correspondingly, the white lines denote the same quantities derived from the orbit superposition. As evident from the middle panes, Model 1 is able to recover not only the mean rotational velocity but also the velocity dispersion as a function of galactocentric distance, including some small details associated with the presence of the bar.}
    \label{fig::vphi_reconstruction}
\end{figure*}

As one can notice in Fig.~\ref{fig::velocity_distributions}, the azimuthal velocity distributions exhibit a more complex behaviour compared to the radial and vertical velocities. Therefore, it is vital to assess how well the azimuthal velocity distribution as a function of distance is recovered. This distribution is also quite interesting because it gives an idea of how well the rotation curve of the galaxy can be reconstructed. Figure ~\ref{fig::vphi_reconstruction} shows the density maps in the $r-\vp$ plane for the snapshot data~(left) and reconstructed densities in Model~1~(middle) and Model~2~(right). The mean azimuthal velocity and $\pm$ velocity dispersion from the snapshot data are shown with cyan lines across the panels. Model 2 shows a number of features non-observed in the snapshot data, where the low-\vp\ secondary overdensity in Fig.~\ref{fig::velocity_distributions}~(right) represents the artificial spheroid whose origin is discussed above. On the contrary, Model 1 shows a very good agreement with the rotational velocity distribution in the snapshot. Some small-scale features can be different; however, this can be explained by the lack of spiral arms in the reconstructed velocity distribution, which, for instance, are responsible for some of the features, e.g. diagonal ridges~\citep{2018Natur.561..360A, 2019MNRAS.490.1026H, 2022A&A...663A..38K}. Another source of discrepancy stems from the fundamental assumption underlying the orbit superposition method: the reliance on the galaxy being in dynamic equilibrium. This assumption is, of course, not exactly correct. For instance, for the orbit integration, we adopted an instantaneous value of the bar pattern speed; however, bars in the MW-mass galaxies are expected to slowdown~\citep{1998ApJ...493L...5D, 2000ApJ...543..704D, 2002ApJ...569L..83A}. On top of this effect, the parameters of bars~(length, strength, pattern speed) oscillate on a short time scale during a connection-disconnection cycle with slowly-rotating spiral arms~\citep{2020MNRAS.497..933H, 2024MNRAS.528.3576V}. Nevertheless, both models capture well the bar OLR-related ridge whose high-\vp\ tail is seen on top of the distribution at $\approx 8$~kpc~\citep{2019MNRAS.488.3324F, 2020MNRAS.494.5936F} suggesting that the instantaneous bar pattern speed value works reasonably well in terms of the large-scale disc kinematics. At the same time, it allows us to study the second-order non-equilibrium effects by subtracting the kinematic orbit superposition solution from the original dataset~(snapshot data or observations).

\begin{figure}
    \centering
    \includegraphics[width=1\hsize]{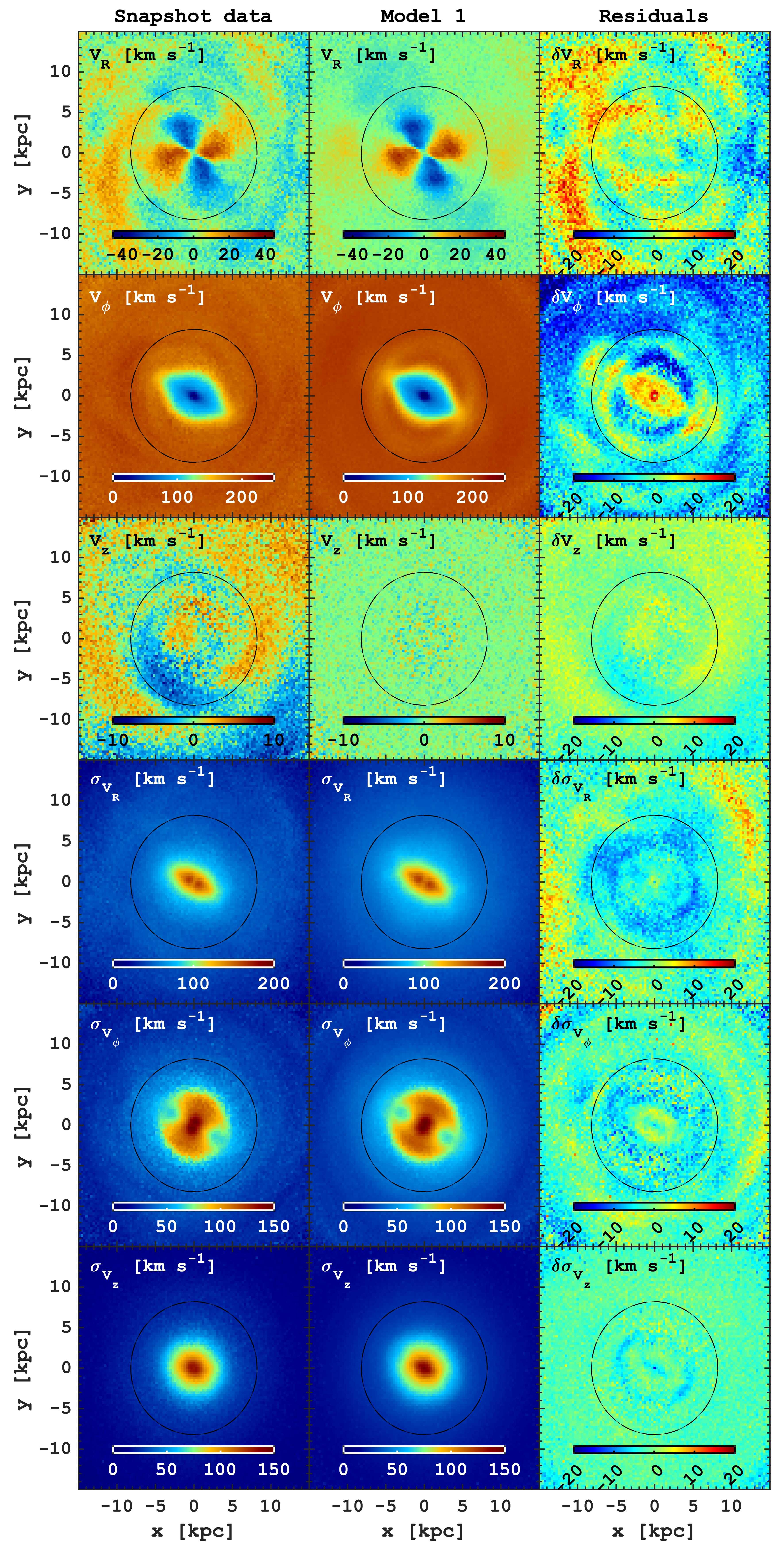}
    \caption{Comparison of the face-on kinematics in the simulation snapshot and orbit superposition Model~1. From top to bottom: the mean velocities \vr, \vp, \vz, and velocity dispersion components \sr, \sphi, \sz. The left column corresponds to the snapshot data, the middle one depicts the orbit superposition, and the right column displays their residuals. The orbit superposition model reproduces kinematic features typical for MW-like barred galaxies, including the quadrupole \vr\ pattern, the axisymmetric rise of the rotational velocity with distance from the centre, \vp, and the 2D velocity dispersion profiles with $\approx 10$~\kmps precision. At the same time, the model is unable to recover the spiral arms-induced velocity perturbations and a weak disequilibrium seen in the \vz\ maps. 
    }
    \label{fig::velocity_maps}
\end{figure}

We conclude this section by presenting the in-plane distribution of the mean and velocity dispersion components. As demonstrated earlier, orbit superposition Model~2 fails to capture several important kinematic characteristics of the simulated galaxy, but it introduces some artificial features. Therefore, in Fig.~\ref{fig::velocity_maps}, we solely compare the snapshot data with velocity distributions from Model~1. 

The radial velocity component of the simulated galaxy exhibits a well-known quadrupole or so-called butterfly pattern in the bar region~\citep{2007MNRAS.379.1155C,2020MNRAS.494.5936F}, a phenomenon that has recently been directly observed in the MW as well~\citep{2019MNRAS.490.4740B, 2023A&A...674A..37G}. Outside the bar region, we see a large-scale radial velocity pattern associated with the presence of a spiral structure in the simulation. The reconstructed in Model~1 radial velocity effectively captures the butterfly pattern. It is worth noting that Mock~1 includes only a small fraction of stars from this inner region (see Fig.~\ref{fig::initial_selection}); however, the superposition of orbits enables us to access the entire pattern. Interestingly, in the recovered velocity distribution outside the bar region, the mean radial velocity is not zero but exhibits  weak yet systematic asymmetric variations still caused by the bar~\citep{2016MNRAS.461.3835M}. As we have shown already, the spiral arms behaviour is not captured by our orbit superposition, which is very well seen in the residual map~(top right panel), depicting the radial velocity pattern outside the bar region. Note that in the case of multi-arm spirals, the radial velocity variations do not trace exactly the density waves.

The main features of the azimuthal velocity distribution~(second row in Fig.~\ref{fig::velocity_maps}) are also precisely reproduced by our model. The bar influences the oval-like shape of the rising inner part of the rotational velocity. Once again, we observe some discrepancies outside the bar region, attributable to the presence of spiral arms, as discussed in the context of the radial velocity distributions. The residual map displays certain features, which we attribute to the precision of the stellar density distribution approximation in the bar region. Although we have demonstrated that the approximated density effectively recovers the snapshot data, minor deviations~(see Fig.~\ref{fig::density_reconstruction_maps}) may lead to the observed small-scale discrepancies in the kinematic data, which, however, for the azimuthal velocity do not exceed 10~\kmps.

\begin{figure*}
    \centering
    \includegraphics[width=1\hsize]{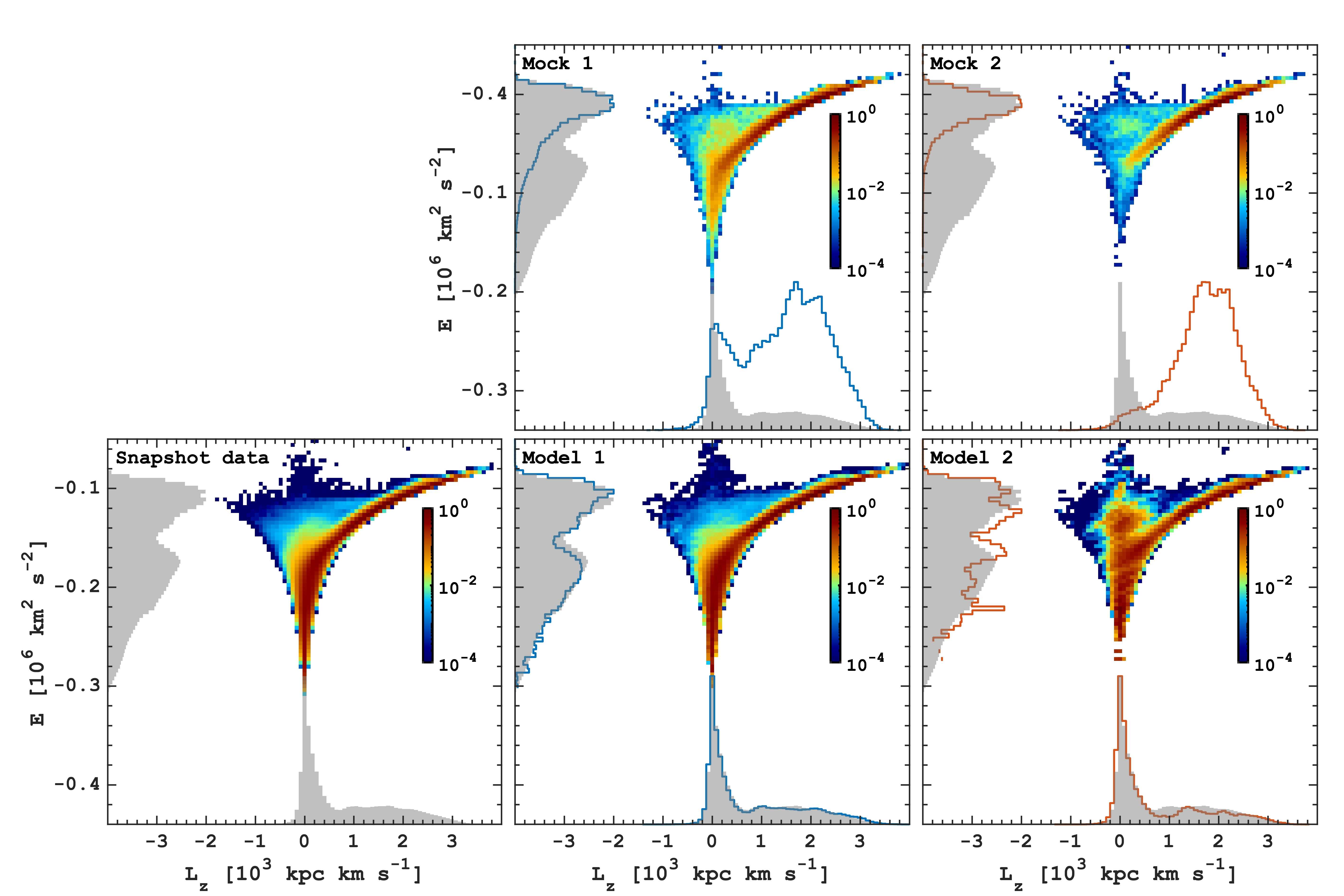}
    \caption{Reconstruction of energy-angular momentum space. The top row displays the initial distribution of stars from the simulation in the \ELz space in two APOGEE-like mock spatial selections adopted for analysis. The bottom row presents a comparison between the snapshot data~(left) and the results of the orbit superposition based on Mock~1~(middle) and Mock~2~(right). Across all panels, the grey-filled area represents the generalized distributions of \E\ and \Lz\ in the snapshot data, contrasted with Mock~1 (blue) and Mock~2 (red) in the middle and models results in the bottom. The density distribution maps are normalized by the maximum value in a given panel. Both initial mocks are biased towards rotationally-supported disc components, as evidenced by their density distributions peaking around the solar radius~(see Fig.~\ref{fig::initial_selection}), thereby diminishing contributions from the innermost galaxy. However, the superposition models in the bottom row effectively correct such selection function biases, accurately reproducing the snapshot data.}
    \label{fig::elz_reconstruction}
\end{figure*}

\begin{figure*}
    \centering
    \includegraphics[width=1\hsize]{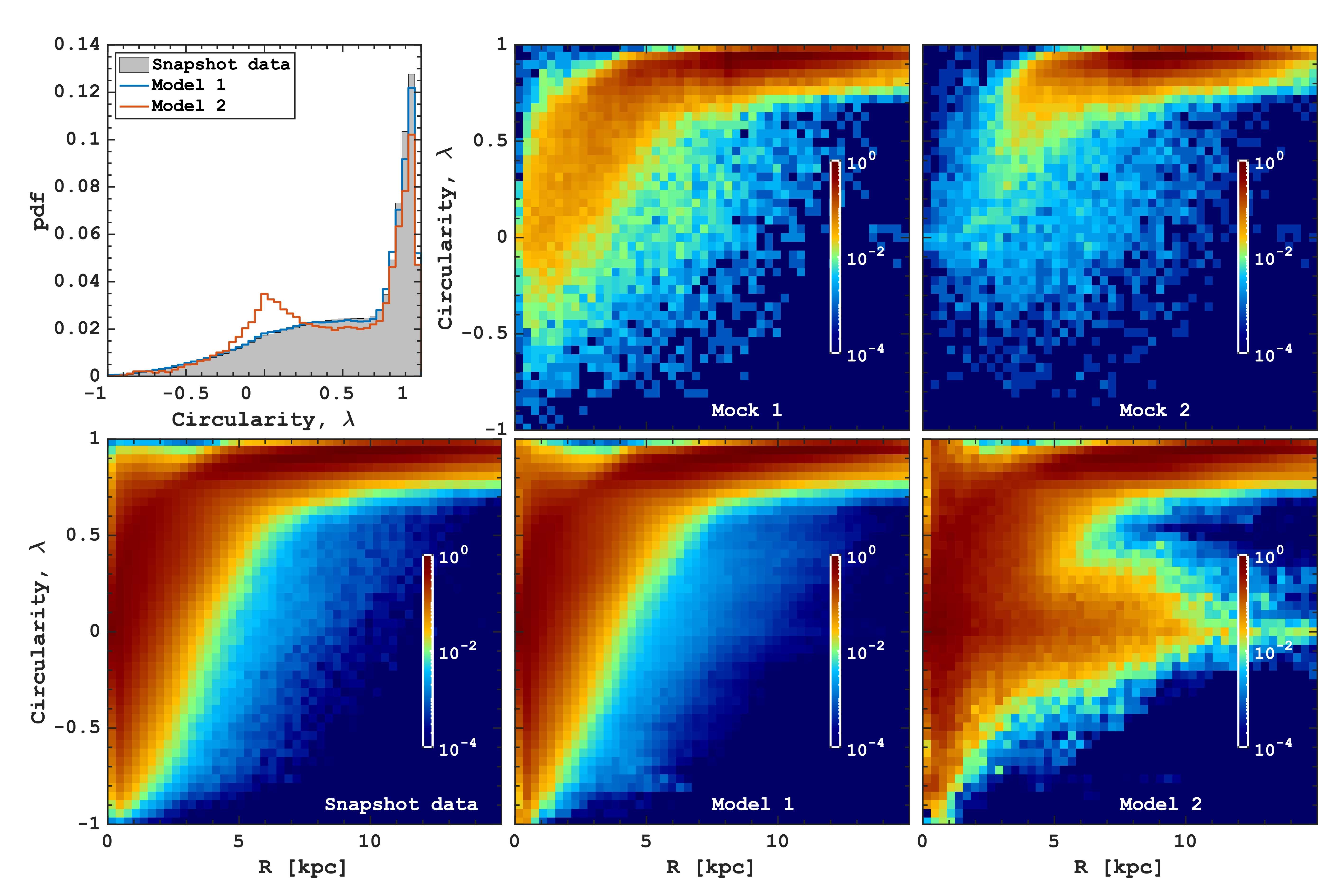}
    \caption{Reconstruction of circularity distribution as a function of the galactocentric distance. Similar to Fig.~\ref{fig::elz_reconstruction}, the panels illustrate the circularity distribution as a function of galactocentric distance in the mock selections~(top) and the results of orbit superposition~(bottom), compared to the snapshot data (bottom left). The top left panel displays the circularity distribution obtained in the orbit superposition (blue and red for Model 1 and 2, respectively) compared to the snapshot data (grey).}
    \label{fig::circularity_reconstruction}
\end{figure*}

Unlike the in-plane velocity components, the reconstructed vertical velocity map exhibits significant discrepancies with the snapshot data (third row in Fig.~\ref{fig::velocity_maps}). The velocity distribution based on Mock~1 appears essentially featureless, as expected from the orbits superposition modelling, which provides an equilibrium solution. However, the simulation snapshot displays some spiral-like vertical velocity corrugations. Since these variations trace the radial velocity pattern associated with spirals, we assume that the vertical velocity features mainly result from the presence of spiral arms, consistent with some previous theoretical works~\citep{2014MNRAS.443L...1D,2014MNRAS.440.2564F,2016MNRAS.457.2569M}; however the effect of the X-shaped bulge formation~\citep{2019A&A...629A..52L,2019A&A...622L...6K} and vertical perturbations excited by the halo particles~\citep{2018MNRAS.480.4244C} might be non-negligible.

Overall, the velocity dispersion components (three bottom rows in Fig.~\ref{fig::velocity_maps}) are recovered quite precisely by our Model~1, exhibiting only minor inconsistencies compared to the snapshot data, typically less than $10$~\kmps. As discussed earlier, these discrepancies likely stem from a slight mismatch of the density~(gravitational potential) approximation and the presence of non-equilibrium kinematic features in the disc, as discussed above.

To summarize the results of this section. We have demonstrated that our orbit superposition approach allows not only the recovery of the adopted 3D stellar density distribution but also the reproduction of the kinematics of stars in great detail across the entire galaxy. The quality of the reconstruction of the kinematic parameters severely depends on the adopted star sample used to calculate the orbital library. However, using mock catalogues reproducing the APOGEE footprint, we demonstrated that the present-day data are already sufficient for such purposes. 

\subsection{Energy, angular momentum and circularity reconstruction}\label{sec::results_elzcirc}
Although we have already explored the quality of the velocity and velocity dispersion components reconstruction by the orbit superposition method, in this section, we aim to address the behaviour of several other parameters often used to explore the MW and external galaxies. In particular, we look at the energy-angular momentum space~(\ELz) and circularity distributions. We emphasize that, unlike axisymmetric potentials, these quantities are not conserved along orbits in a steadily rotating barred potential~\citep{2007MNRAS.379.1155C, 2008gady.book.....B}. Hence, a single orbit can contribute with different energy values, angular momentum and circularity. However, it is important to recall that the weight attached to different values of kinematic parameters along a single orbit remains constant throughout.

In Fig.~\ref{fig::elz_reconstruction} we show the stellar density distribution in the \ELz\ coordinates. The top panels present the distributions of the input Mock~1 and Mock~2 star particles selection, which do not cover the whole disc~(see the spatial footprints in Fig.~\ref{fig::initial_selection}) but are somewhat weighted to the solar radius, which is evident from the panels, showing the dominance of kinematically-cold discy component. These APOGEE-like footprints capture relatively few low-energy and low-angular momentum stars, which can be found throughout the entire disc but may not necessarily represent the genuine innermost region of the galaxy. Interestingly, both mock catalogues (in Mock~2, it is more prominent, top right panel) reveal a blob at high energy values and weak-to-no rotation separated from the low-energy tail by a gap. As it is known in the literature~\citep{2022MNRAS.510.5119L,2022MNRAS.514..689B}, and highlighted by Fig.~\ref{fig::elz_reconstruction}, this \ELz\ distribution structure is solely caused by the APOGEE footprint. We recall that our simulated galaxy evolved in isolation and, thus, does not include accreted populations and is obviously not affected by external interactions, highlighting the importance of critical analysis of the spatial selection function impact on the \ELz\ space~(and other kinematic or actions coordinates) often used for the kinematic substructures identification and merger debris analysis in the MW halo~\citep{2017A&A...604A.106J, 2023A&A...677A..91K, 2023A&A...673A..86P, 2024A&A...690A.136M}.

The snapshot data in the \ELz\ plane (bottom left panel in Fig.~\ref{fig::elz_reconstruction}) reveals at least one striking feature associated with the presence of the bar. Several horizontally aligned overdensities are visible, representing the resonances of the bar in the stellar disc~\citep{2023MNRAS.524.3596D,2024arXiv240214907D}. The signal observed in our disc-galaxy simulation is much stronger compared to the one found by \cite{2023MNRAS.524.3596D}; however, one can expect some contamination from disc stars in the halo. This is partly because these two components have no sharp boundaries, especially if the inner MW halo is composed predominantly of heated disc stars~\citep{2019A&A...632A...4D,2020MNRAS.494.3880B}.

Once we compared the reconstructed stellar density distribution with the snapshot data, we observed that Model~1 reproduces very well the distribution of \E, \Lz\ separately, as well as the detailed structure of the \ELz\ space. The model captures nearly constant energy overdensities, highlighting their bar-related origin. The Model~2 also closely resembles the true distributions, recovering the \Lz\ distribution well but exhibiting less accuracy in the \E\ distribution. The latter displays several peaks and multiple artificial features in \ELz\ coordinates, likely resulting from a limited and possibly not fully representative orbital library. In this situation, the orbit superposition adds too much weight to certain existing orbit families while many important ones are missing, as we already discussed in this and previous sections.  Therefore, caution should be exercised when interpreting features in the \ELz\ analysis based on the orbit superposition models, as the outcome of such models should be thoughtfully tested depending on the input sample of stars reflecting the survey selection function.

Another parameter widely used to explore both structural and kinematic structure of galaxies is orbital circularity, the ratio of stellar specific angular momentum and the angular momentum of a perfectly circular orbit with the same energy~\citep{2003ApJ...597...21A,2004ApJ...612..894B,2012MNRAS.420..255T}, which also plays a vital role in the interpretation of orbit superposition models of external galaxies~\citep{2020MNRAS.491.1690J, 2018MNRAS.473.3000Z, 2018NatAs...2..233Z}. In Fig.~\ref{fig::circularity_reconstruction}, we depict the circularity distribution as a function of galactocentric distance for the mock selections of star particles~(top middle and right), snapshot data (bottom left), and reconstructed distributions in Model~1 and Model~2~(bottom middle and right). The top left panel displays the circularity distribution in the snapshot data~(grey) and the distributions obtained using orbit superposition in Model~1 and Model~2, depicted in blue and red, respectively. As we showed earlier, here, Model~1 works perfectly and recovers not only the overall distribution of circularity but also its variation with the galactocentric distance. The fraction of hot orbits dominating the inner galaxy and the transition to colder disc-like populations are very much identical to the snapshot data. Conversely, Model~2 exhibits an artificial overabundance of hot orbit stars with circularity values near zero. This discrepancy was previously observed and discussed in detail above~(see Figs.~\ref{fig::vphi_reconstruction} and ~\ref{fig::elz_reconstruction}). 


\subsection{Reconstruction of metallicity distribution}\label{sec::results_feh}
The ultimate stress test for our orbit superposition method, based on the MW-like data, lies in understanding the reconstruction of stellar parameter variations across the galaxy. It is important to note that no stellar parameters~(chemical abundances and stellar ages) were used in the modelling process or generation of the mock catalogues. However, given the evolutionary coupling between chemical abundances and orbits of stars~\citep{2003A&A...410..527B, 2008MNRAS.388.1175H, 2009A&A...501..941H, 2012ApJ...755..115B, 2013A&A...560A.109H, 2014ApJ...781L..20M, 2019ApJ...883..177N}, we anticipate that our model can also reasonably capture the metallicity structure of the simulated galaxy. We remind the reader that here, we use an isolated galaxy model where the enrichment history and, thus, the metallicity scale may not be representative of the MW.

\begin{figure}
    \centering
    \includegraphics[width=1\hsize]{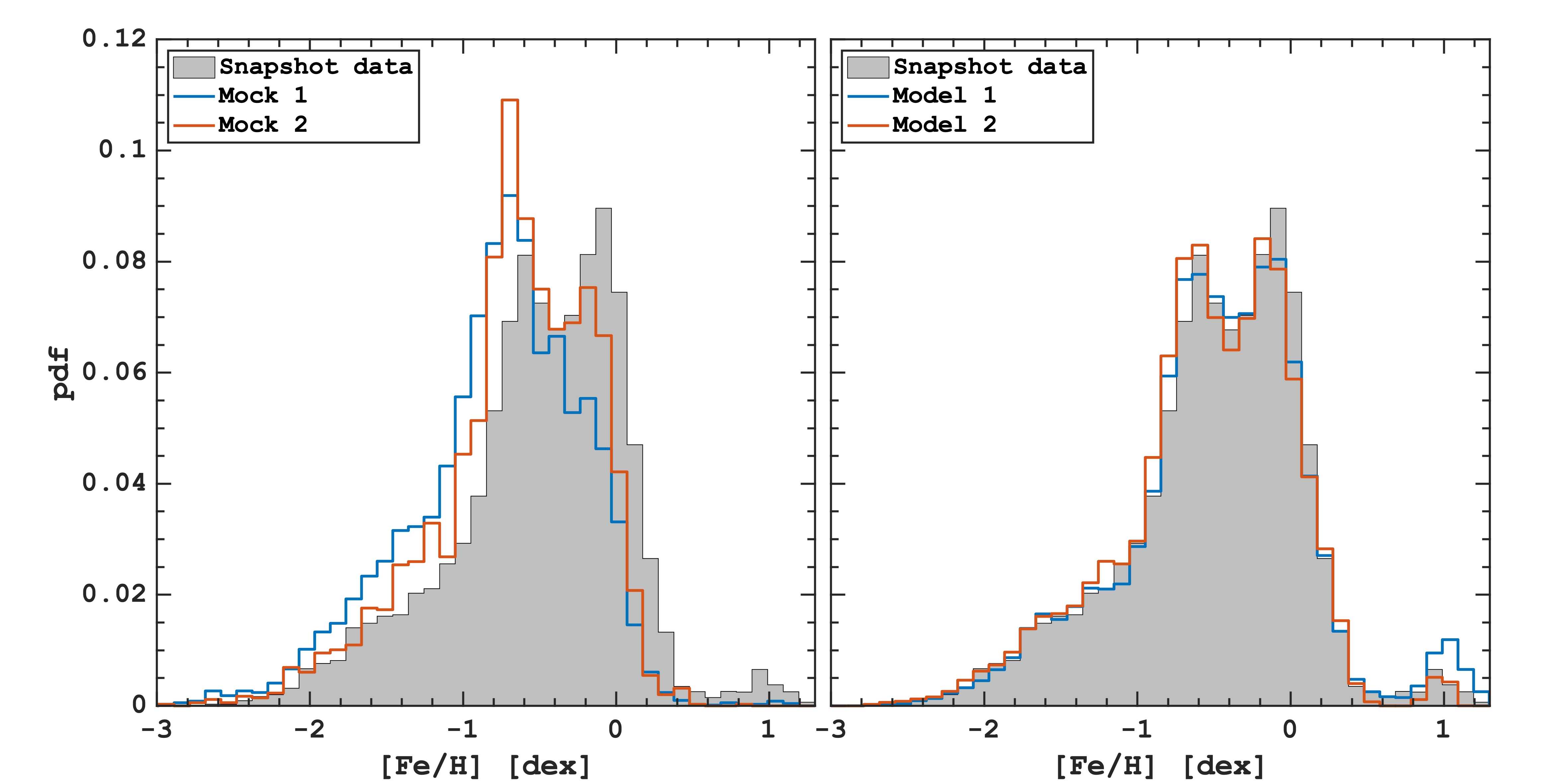}
    \caption{The left panel displays the \FeH distribution for stars in the initial model selections (blue and red). The right panel shows the results of the orbital decomposition. In both panels, the MDF of stars in the simulation snapshot is represented by the grey-filled histogram. The MDFs of the initial selections are biased against relatively metal-rich stars, which is corrected by the orbit superposition modelling. We emphasize that our approach enables the recovery of the complex shape of the MDF, characterized by multiple peaks, including a small fraction of extremely metal-rich stars originating from the centre of the simulated galaxy. This underscores the capability of our method to reconstruct any characteristics~(abundances and ages) of stellar populations that are evolutionarily linked to the orbital parameters of stars.}
    \label{fig::MDF_reconstruction}
\end{figure}

First, we access the reconstruction of the MDF of the entire galaxy. In Fig.~\ref{fig::MDF_reconstruction} we show the MDF of the simulated snapshot~(grey) considering all stars formed in the simulation and located within $15$~kpc. The Mock 1 and Mock 2 selections are shown with blue and red histograms in the left panels, revealing a bias towards lower metallicity stars, as is expected from their spatial footprints, which do not capture many relatively metal-rich stars in the inner galaxy. Nevertheless, both initial selections still cover the full range of metallicity, yet the fractional contributions of stars in both samples do not reproduce the true MDF. The full range of metallicity in the initial sample is crucial to recovering the true MDF, again highlighting the importance of the initial sample of stars and the need for the complete coverage of the parameter space. The right panel shows the orbit superposition-based MDFs~(Model 1 and Model 2) against the simulation snapshot data. We notice that both models recover the true MDF quite well. They recover all three peaks of the MDF, even the most metal-rich one, which was poorly populated in the corresponding mocks. The figure demonstrates the advantage of the orbit superposition methods, as it allows to rebalance the metallicity~(or any other stellar parameters) distribution in a given spectroscopic dataset, thus providing selection-function-corrected distributions.

\begin{figure*}
    \centering
    \includegraphics[width=1\hsize]{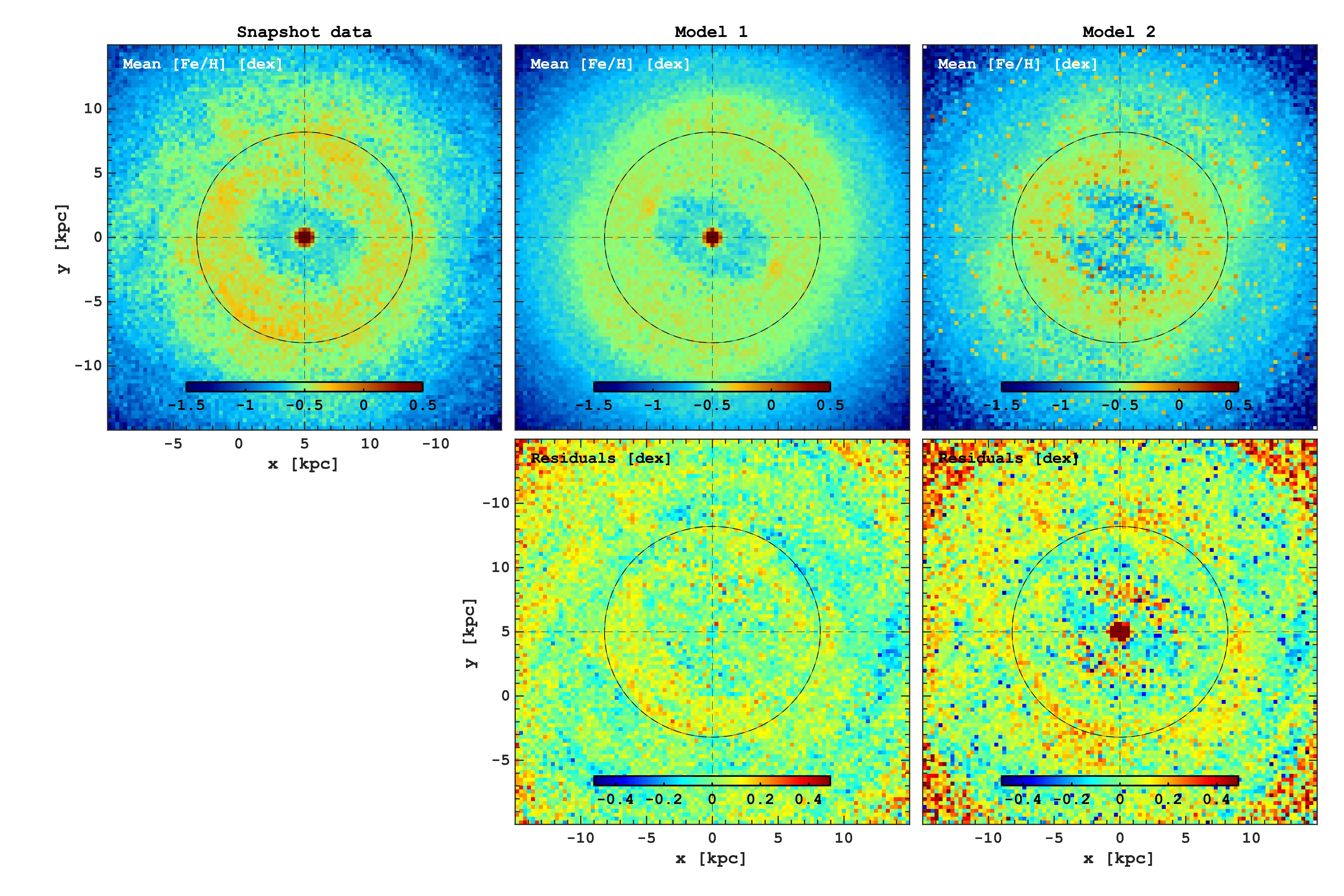}
    \caption{Reconstruction of the face-on stellar disc metallicity maps. The top panels display the mean \FeH\ maps for the snapshot data (left) and the orbit superposition modelling based on Model~1~(middle) and Model~2~(right). The bottom panels illustrate the residuals between the snapshot data and the reconstructed metallicity maps. While the models effectively reproduce the overall 2D behaviour of metallicity, they fail to capture the enhancement of the mean metallicity associated with spiral arms.}
    \label{fig::metallicity_maps_reconstruction}
\end{figure*}

Finally, we compare the 2D face-on maps of the mean stellar metallicity obtained in our orbit superposition models. Figure \ref{fig::metallicity_maps_reconstruction} shows the comparison between the true~(simulation snapshot data) metallicity distribution~(top left), reconstructed maps based on orbit superposition~(top middle and right) and corresponding residuals in the bottom panels. We notice that the metallicity distribution in the snapshot data is quite complex. Again, it might not be typical for the disc galaxies and the MW in particular; however, it allows us to test the ability of our approach to reconstruct the complex abundance distributions. Nevertheless, some elements are consistent with our expectations about the metallicity distribution across discs of barred galaxies. The metallicity peaks in the very compact region representing a metal-rich core likely associated with efficient enrichment in the region fuelled by the gas inflow along the bar~\citep{2024A&A...690A.352R} and recently explored in the MW~\citep{2024arXiv240601706R}. Inside the bar region, the metallicity is lower with a modest enhancement along the bar~\citep{2013A&A...553A.102D, 2018A&A...616A.180F, 2024MNRAS.534.2438N}. Outside the bar region, the metallicity is higher, and it decreases towards the disc outskirts. The spiral arms are seen as having slightly higher metallicity compared to the surrounding area, as expected in various scenarios~\citep{2016ApJ...830L..40S,  2019A&A...628A..38S, 2018A&A...611L...2K, 2023A&A...671A..56K}.

Model~1 captures most of the elements of the metallicity map we have in the simulation: the central metallicity peak, lower metallicity region inside the bar region and a decline of the mean stellar metallicity outside the bar. We also note that there is a prominent large-scale azimuthal variation of the mean metallicity outside the bar, which is not associated with spiral arms and seems to be observed in the MW~\citep{2023MNRAS.525.3318H}, as we suggest can be caused by the bar. The spiral arms-related increase of the metallicity is not seen in the reconstructed metallicity map based on Model~1; however, it is prominent in the residuals. Model 2 surprisingly well recovers these features. However, the solution is quite noisy, the metal-rich core is completely missing, and the residuals show more peculiar features. Nevertheless, the overall metallicity map in Model~2 is quite adequate, considering how poorly it recovers the density and kinematic information.

\section{Discussion}\label{sec::method_discussion}

\subsection{Limitations of the orbit superposition approach}

While we have demonstrated the robustness of our approach in mapping the galactic disc using the orbit superposition, we need to acknowledge the limitations inherent in the method. The method we describe aims to reconstruct the present-day complete distribution function of the galaxy using patchy information only about the distribution of its resolved stellar populations, and formally, it does not provide or use any information about the galactic structure in the past. The orbital library we construct is obtained in a rotating reference frame, where the time variable is essentially excluded. Although we have created a galaxy model implementing the rotating bar, axisymmetric models can be useful for various applications, such as the halo structure reconstruction, where the impact of the bar and other disc asymmetries are not dominant.

As discussed in the relevant sections, we underline that our model can not capture the non-equilibrium kinematic features naturally present in galactic discs. Spiral arms, in principle, can be included in the potential reconstruction and orbital library calculation; however, this would imply that spirals are steady structures that maintain their strength and pattern speed on a long time scale. This is quite a very strong requirement and is not supported by the modern literature~\citep{2014PASA...31...35D, 2022ARA&A..60...73S}, suggesting that spirals are transient, recurrent and initiated by swing amplified instabilities. A constant speed of the bar is another assumption which might not perfectly represent the situation in barred galaxies. As we discussed above, non-equilibrium potential, such as a rapidly evolving (buckling) bulge or bar slowing down or accelerating, cannot be adequately captured. The latter, however, can be implemented but requires a creative approach for projecting the orbits of stars into the present-day stellar populations allowing to avoid phase-mixing.

It is well-known that the outer disc of the MW is warped, which is seen not only from the star count~\citep{2001ApJ...556..181D,2002A&A...394..883L} but also in the kinematics of stars~\citep{2018MNRAS.481L..21P,2018MNRAS.478.3809S}. Our model is not able to reproduce such a phenomenon unless the approximated gravitational potential includes such a feature. This can be implemented; however, this would require a steady rotation of the warp or an assumption about its evolution in the past, which might not be so trivial taking into account its precession~\citep{2020NatAs...4..590P,2024arXiv240509624H}. 

The results of the orbit superposition are governed by the adopted potential of the galaxy, which is trivial to obtain using a simulated galaxy; however, in practice, the true potential of the MW is not known. Therefore, although several models of the MW potential are available~\citep{2015ApJS..216...29B, 2017MNRAS.465...76M, 2017A&A...598A..66P, 2017MNRAS.465.1621P, 2022MNRAS.514L...1S}, great caution should be addressed to the potential adopted in the context of the real MW data. In this case, the pattern speed and the present-day orientation of the MW bar can play a vital role, too, which is not a major problem in the simulated data. At the same time, a non-reliable potential is likely to result in odd kinematic features compared to the real data, and we suggest that at least the rotational velocity should be well-recovered by the orbit superposition.

In this work, we showed that two different selections of star particles can result in different solutions. This aligns with the results from Schwarzschild modelling of external galaxies, which underscore the significance of large-area coverage and high spatial resolution constraints offered by IFU data compared to the long-slit observations ~\citep{2005CQGra..22S.347C}. The enhanced capabilities of IFU data make it possible to resolve degeneracies in determining the galaxy's gravitational potential, including more precise constraints on the parameters of the central SMBHs. To obtain more robust results in the context of the MW, the sample of stars~(providing orbits used for superposition) should at least partially include a contribution of stars originating from the inner galaxy and whose apocenters lie inside the initial sample area~(mock catalogue). This is particularly important for the reconstruction of the galactic nuclei and the bulge region. Similarly, the orbits of those stars in the outer disc whose pericenters fall outside the selection footprint will not be accounted for in the model, which will be unrealistic at the outskirts.

The selected sample of stars also should cover the whole range of stellar parameters if one is interested in mapping metallicities, elemental abundances and ages. The orbit superposition models developed here do not provide additional information about the stellar populations but rather allow us to extrapolate or project their parameters along their orbital space. Therefore, the model may not adequately represent reality if the abundance or age information in the input sample of stars is incomplete or heavily biased. In any case, the ground truth about the MW stellar populations is unknown as different methods can deliver different abundances from the same stellar spectra~\citep[see, e.g.][]{2022ApJS..262...34H} and various age determination methods, although they generally agree with each other, may provide different ages for the same stars~\citep[see, e.g.][]{2023A&A...678A.158A, 2023A&A...673A.155Q}. Therefore, the orbit superposition models should be considered as a projection or extension of a given set of parameters to the whole galaxy under certain assumptions regarding its gravitational potential and orbital library.

\subsection{Future prospects}

It is important to recognize that the limitations discussed in the previous section allow us to investigate processes influenced by factors not included in our model. For instance, we have shown the metallicity and velocity patterns associated with spiral arms are quite prominent in the residual maps~(see Fig.~\ref{fig::metallicity_maps_reconstruction} and~\ref{fig::velocity_maps}). Thus, such systematic variation of stellar parameters is seen once the data are compared with the outcome of the orbit superposition models. This allows us to subtract the background information~(orbit superposition results) from the data and focus on specific phenomena, like spirals and the warp. 

The origin of the galactic spiral structure remains largely elusive; hence, testing what we can learn from models with a steady or evolving spiral structure can be very promising. By experimenting with the strength and the pattern speed of spirals, one can find to what extent the steady spiral structure is not applicable for recovering the observed data. Since its discovery~\citep{1991ApJ...379..631B,1991Natur.353..140N,1992ApJ...384...81W}, a lot of efforts has been paid to pin down the parameters of the MW bar~\citep{2016ARA&A..54..529B}. A similar strategy we propose for spiral arms study can be pursued for finding the bar parameters if one can vary the structure and kinematics of the bar and use some known parameters of stellar populations, e.g. kinematics or abundance distribution, to find the most appropriate solution. For instance, one can try to constrain the bar parameters by matching its resonance `location' in various kinematic spaces. In some sense, such an approach would be similar to the made-to-measure~(M2M) models~\citep{2007MNRAS.376...71D, 2009MNRAS.395.1079D, 2013MNRAS.430.1928H, 1996MNRAS.282..223S, 2010MNRAS.405..301L}. However, the main advantage compared to the existing M2M models is that in the orbit superposition approach, one can also use the stellar parameters information, thus putting more constraints on the model and learning more about the present-day structure of the MW stellar populations.

Finally, the implemented orbit superposition model is relatively simple, as it assumes that the total and stellar disc potentials (or density distributions) are known, allowing us to bypass the use of kinematic data. The latter statement, however, may not be entirely precise if the method is applied to the MW because the parameters of its main components~(DM, stellar disc and gas density distributions) are, in fact, constrained using the kinematics of various tracers which are not part of the orbit superposition modelling. Nevertheless, despite progress in constraining the 3D density distribution of the MW, as discussed in the paper, some uncertainties remain. Therefore, we propose that future developments of orbit superposition models for the MW should relax the requirement for a known galactic potential. Instead, the potential could be constrained by incorporating kinematic information (higher-order moments of the distribution function) from the initial sample of stars used to create the orbital library. This step is not trivial, as it would require a convolution of the orbit superposition model results with observational errors (uncertainties in radial velocity, distance, and proper motions) and their 're-observation' using the same spatial footprint as the input dataset. This formalism then will be identical to the modern Schwarzschild orbit superposition models currently developed only for external galaxies.

\section{Summary}\label{sec::method_summary}

In this paper, we described a novel orbit superposition method in the context of the MW data about its resolved stellar populations. The method was extensively tested on a simulated MW-like barred galaxy and can be outlined in the following steps.
\begin{itemize}
    \item[1.] Define the total galactic gravitational potential and the three-dimensional stellar mass density distribution component.
    \item[2.] Integrate the orbits of stars from the observational sample within the total potential.
    \item[3.] Project the integrated orbits and the 3D stellar density onto a grid.
    \item[4.] Calculate non-negative weights for each orbit to ensure their superposition in each grid cell corresponds to the adopted stellar mass.
\end{itemize}
The resulting orbital weights represent the stellar mass-weighed complete distribution function.

The main results of the paper are as follows.
\begin{itemize}
    \item To test the impact of the initial sample of stars on the orbit superposition model results, we used two mock catalogues generated from a simulated MW-like barred galaxy. These catalogues were designed to mimic the spatial distribution of giant stars~(Mock 1) and red clump stars~(Mock 2) in the APOGEE DR17 dataset (see Fig.~\ref{fig::initial_selection}).

    \item We demonstrated that orbit superposition Model~1~(based on Mock 1) shows remarkable agreement with the simulated galaxy in terms of the mean velocity and velocity dispersion components, as well as circularity, energy, and angular momentum distributions (see Sections \ref{sec::results_density} - \ref{sec::results_elzcirc}).
    
    \item Model 2 fails to reproduce certain details of the velocity distributions accurately, but it introduces several artificial density and velocity features, particularly in the inner region of the galaxy. As a result, the corresponding spatial footprint~(Mock 2) is less suitable for applications related to the MW. The divergence in results between the considered models stems from the incomplete orbital families in Model 2. This model input mock data lacks a sufficient number of stars with apocentres in the innermost region of the galaxy, instead including only hot orbit stars that pass through the initial footprint and, as a result, the orbits of these stars receive too much weight and produce an artificial spheroid-like stellar component, not present in the simulation.

    \item We showed that the orbit superposition approach allows the recovery of the MDF of the entire galaxy and the face-on mean stellar metallicity distributions. This showcases the ability of our approach to correct observed stellar parameters distributions and kinematic structure for selection functions of Galactic surveys and deliver reliable information about the chemo-kinematic properties of stellar populations even outside the survey spatial footprint. 
    
    \item We have shown that the present-day data, in particular, APOGEE DR 17, is already sufficient to produce a high-quality orbit superposition analysis of the MW galaxy. However, the number of stars with precise kinematic and stellar parameter measurements is poised for a significant increase with the upcoming release of data from surveys such as 4MOST, SDSS-V, MOONS, and WEAVE.  Therefore, the proposed orbit superposition approach and its further extensions can then be used to project these data and reveal the MW stellar populations in great detail.
    
\end{itemize}

The newly developed approach potentially allows the recovery of an unbiased present-day age, chemical abundance and kinematic structure of the MW galaxy, which can then be used to reconstruct the formation of its main components and their evolutionary connection. The application of the developed method to the real MW data will be presented in the follow-up papers.

\begin{acknowledgements}
SK thanks Vasily Belokurov for a discussion about the impact of the APOGEE footprint on the distribution function and James Binney for some useful suggestions regarding the application of the method.

\thanksleiden\
\\
\thanksmiapb\
\\
\thankssdss\
\\
\thanksgaia
\\
\end{acknowledgements}

\bibliographystyle{aa}
\bibliography{refs}

\end{document}